\documentclass[acmlarge,nonacm]{acmart}

\newcommand{\ignore}[1]{ }
\usepackage{graphicx}
\usepackage{textcomp}

\AtBeginDocument{%
  \providecommand\BibTeX{{%
    \normalfont B\kern-0.5em{\scshape i\kern-0.25em b}\kern-0.8em\TeX}}}

\setcopyright{none}

\def\mode{0}  

\def\redactionmode{1} 

\def\alttextmode{1} 
\usepackage{amsmath}
\usepackage{color}
\usepackage{cleveref}
\usepackage{comment}  	

\usepackage{enumitem}				
\usepackage{graphicx}
	\graphicspath{{./figures/}}		
\usepackage{multirow}				

\usepackage{tabularx} 				
\usepackage{threeparttable}
\usepackage{subfigure}
\usepackage{soul} 					
\usepackage{xargs}    				
\usepackage{xcolor}			

\newcommand{\red}{}
\newcommand{\green}{}
\newcommand{\blue}{}

\newcounter{documentmode}
\includecomment{conf}  	

\definecolor{darkgreen}{RGB}{34,139,34}

\newcommand{\setdocmode}{%
  \ifcase\number\value{documentmode}
	\renewcommand{\red}{\textcolor{red}}
	\renewcommand{\green}{\textcolor{darkgreen}}
	\renewcommand{\blue}{\textcolor{blue}}
  \or
    \usepackage{draftwatermark}
	\SetWatermarkScale{1.35}
	\SetWatermarkAngle{56}
	\SetWatermarkLightness{0.9}
	\SetWatermarkHorCenter{0.9\textwidth}
	\SetWatermarkVerCenter{0.7\textheight}
  \or
	\usepackage{draftwatermark}
	\SetWatermarkText{CONFIDENTIAL}
	\SetWatermarkScale{0.67}
	\SetWatermarkAngle{56}
	\SetWatermarkLightness{0.9}
	\SetWatermarkHorCenter{0.6\textwidth}
	\SetWatermarkVerCenter{0.6\textheight}
  \else
    \excludecomment{conf}
  \fi
}

\if\mode0
\usepackage[prependcaption, textsize=footnotesize]{todonotes}
\else
\usepackage[disable, prependcaption, textsize=footnotesize]{todonotes}

\fi

\setdocmode

\newcounter{redmode}
\includecomment{2Detailed}			
\includecomment{Uncritical}			
\includecomment{2SaveSpace}			
\includecomment{IfMustDelete}			
\newcommand{\redact}{%
  \ifcase\number\value{redmode}
    \includecomment{2Detailed}
    \includecomment{Uncritical}
    \includecomment{2SaveSpace}
    \includecomment{IfMustDelete}
  \or
    \excludecomment{2Detailed}
    \includecomment{Uncritical}
    \includecomment{2SaveSpace}
    \includecomment{IfMustDelete}
  \or
    \excludecomment{2Detailed}
    \excludecomment{Uncritical}
    \includecomment{2SaveSpace}
    \includecomment{IfMustDelete}
  \or
    \excludecomment{2Detailed}
    \excludecomment{Uncritical}
    \excludecomment{2SaveSpace}
    \includecomment{IfMustDelete}
  \or
    \excludecomment{2Detailed}
    \excludecomment{Uncritical}
    \excludecomment{2SaveSpace}
    \excludecomment{IfMustDelete}
  \fi
}
\redact

\newcounter{alttxtmode}
\includecomment{TextA}			
\includecomment{TextB}			
\newcommand{\alttext}{%
  \ifcase\number\value{alttxtmode}
    \includecomment{TextA}
    \includecomment{TextB}
  \or
    \excludecomment{TextA}
    \includecomment{TextB}
  \or
    \includecomment{TextA}
    \excludecomment{TextB}
  \or
    \excludecomment{TextA}
    \excludecomment{TextB}
  \fi
}
\alttext

\usepackage[acronym]{glossaries} 

\newacronym{asd}{ASD}{Autism Spectrum Disorder}
\newacronym{cart}{CART}{Classification And Regression Tree}
\newacronym{dfa}{DFA}{Discriminant Function Analysis} 
\newacronym{ecg}{ECG}{Electrocardiography}
\newacronym{skin_eda}{EDA}{Electrodermal activity} 		
\newacronym{ews}{EWS}{Early Warning Score}
\newacronym{ga}{GA}{Genetic Algorithm}
\newacronym{hr}{HR}{Heart Rate}
\newacronym{ldf}{LDF}{Linear Discriminant Function}
\newacronym{rmsdd}{RMSDD}{Root-Mean Square of Successive Differences}
\newacronym{rsvm}{RSVM}{Reputation-driven Support Vector Machine }
\newacronym{saews}{SA-EWS}{Self-Aware Early Warning Score}
\newacronym{sam}{SAM}{Self-Assessment Manikin} 
\newacronym{selphys}{SelPhyS}{Self-aware cyber-Physical System}
\newacronym{sdnn}{SDNN}{Standard Deviation Normal-to-Normal-Intervals}
\newacronym{skt}{SKT}{Skin Temperature}
\newacronym{som}{SOM}{Self-Organizing Map}
\newacronym{whs}{WHS}{Wearable Health-care Systems}

\newacronym{ai}{AI}{Artificial Intelligence}
\newacronym{ann}{ANN}{Artificial Neural Network}
\newacronym{bpn}{BPN}{Back-Propagation Neural Network}
\newacronym{bpnn}{BPNN}{back-propagation neural network}
\newacronym{cnn}{CNN}{Convolutional Neural Network}
\newacronym{flop}{FLOP}{Floating Point Operation}
\newacronym{icnn}{ICNN}{Iterative Convolutional Neural Network}
\newacronym{svm}{SVM}{Support Vector Machine}
\newacronym{nn}{NN}{Neural Network}
\newacronym{nb}{NB}{Naive Bayesian}
\newacronym{mcs}{MCS}{Multiple Classifier System}
\newacronym{ml}{ML}{Machine Learning}
\newacronym{ucnn}{$\mu$CNN}{Micro CNN}

\newacronym{brs}{BRS}{Bipolar Resistive Switch-based logic}
\newacronym{cnf}{CNF}{Conjunctive Normal Form}
\newacronym{crs}{CRS}{Complementary Resistive Switch-based logic}
\newacronym{dnf}{DNF}{Disjunctive Normal Form}
\newacronym{fpm}{FPM}{Forward Polarized Memristor}
\newacronym{hfo}{$HfO_x$}{Hafnium Oxide}
\newacronym{hrs}{HRS}{High Resistance State}
\newacronym{imc}{IMC}{In-Memory Computation}
\newacronym{imply}{IMPLY}{Material Implication}
\newacronym{imp}{IMP}{In-Memory Processing}
\newacronym{lim}{LIM}{Logic in Memory}
\newacronym{lrs}{LRS}{Low Resistance State}
\newacronym{ltg}{LTG}{Logic Threshold Gate}
\newacronym{magic}{MAGIC}{Memristor-Aided Logic}
\newacronym{mecoins}{Me-Coin}{Memristor-based Computation In-memory}
\newacronym{pcm}{PCM}{Phase Change Memory}
\newacronym{pim}{PIM}{Processing in Memory}
\newacronym{reram}{ReRAM}{Resistive Random Access Memory}
\newacronym{rpm}{RPM}{Reversely Polarized Memristor}
\newacronym{stt}{STT}{Spin Transfer Torque}
\newacronym{tao}{$TaO_x$}{Tantalum Oxide}
\newacronym{tio}{$TiO_2$}{Titanium dioxide}
\newacronym{vteam}{VTEAM}{Voltage-controlled ThrEshold Adaptive Memristor}

\newacronym{aco}{ACO}{Autonomous Cooperating Object}
\newacronym{afdd}{AFDD}{Automated Fault Detection and Diagnostic}
\newacronym{ca}{CA}{Continuous Average}
\newacronym{cah}{CAH}{Context-Aware Health Monitoring}
\newacronym{cam}{CCAM}{Confidence-based Context-Aware condition Monitoring}
\newacronym[plural=DABs,longplural={Discrete Average Blocks}]{dab}{DAB}{Discrete Average Block}
\newacronym[plural=KPNs,longplural={Kahn Process Networks}]{kpn}{KPN}{Kahn Process Networks} 
\newacronym{mape-k}{MAPE-K}{Monitor-Analyze-Plan-Execute over a shared Knowledge}
\newacronym[plural=MoCs,longplural={Models of Computation}]{moc}{MoC}{Model of Computation}
\newacronym{oda}{ODA}{Observe-Decide-Act}
\newacronym{pca}{PCA}{Principal Component Analysis}
\newacronym{rosa}{RoSA}{Research on Self-Awareness}
\newacronym{sa}{SA}{Self-Aware}
\newacronym{saness}{SA}{Self-Awareness}
\newacronym{sahm}{SAHM}{Self-Aware Health Monitoring}
\newacronym{samba}{SAMBA}{Self-Aware health Monitoring and Bio-inspired coordination for distributed Automation systems}
\newacronym{sh}{SH}{State Handler}

\newacronym{cps}{CPS}{Cyber-Physical System}
\newacronym{cpps}{CPPS}{Cyber-Physical Production System}
\newacronym{dsr}{DSR}{Down-Sampling Rate}
\newacronym{dum}{DuM}{Device under Monitoring}
\newacronym{es}{ES}{Embedded System}
\newacronym{mes}{MES}{Manufacturing Execution System}
\newacronym{sc}{SC}{Stochastic Computing}
\newacronym{sos}{SoS}{System of Systems}
\newacronym{suo}{SuO}{System under Observation}

\newacronym{abi}{ABI}{Application Binary Interface}
\newacronym{adc}{ADC}{Analog-to-Digital Converter}
\newacronym{aes}{AES}{Advanced Encryption Standard}
\newacronym{alu}{ALU}{Arithmetic Logic Unit}
\newacronym{api}{API}{Application Programming Interface}
\newacronym{asic}{ASIC}{Application Specific Integrated Circuit}
\newacronym{asoc}{ASOC}{Autonomic System-on-Chip platform}
\newacronym{axi}{AXI}{Advanced eXtensible Interface Bus}
\newacronym{bram}{BRAM}{Block Random Access Memory}
\newacronym{cdt}{CDT}{C/C++ Development Tooling}
\newacronym{clb}{CLB}{Configuarable Logic Block}
\newacronym{cmos}{CMOS}{Complementary Metal-Oxide Semiconductor}
\newacronym{cp}{CP}{Clock Pulse}
\newacronym{cpi}{CPI}{Cycles Per Instruction}
\newacronym{cpu}{CPU}{Central Processing Unit} 
\newacronym{cpsoc}{CPSoC}{Cyber-Physical System-on-Chip}
\newacronym{cu}{CU}{Compute Unit}
\newacronym{cuda}{CUDA}{Compute Unified Device Architecture}
\newacronym{dac}{DAC}{Digital to Analog Converter}
\newacronym{ddr3}{DDR3}{Double Data Rate}
\newacronym{dff}{DFF}{Data Flip-Flop}
\newacronym{dll}{DLL}{Delay Locked Loop}
\newacronym{dmr}{DMR}{Dual Modular Redundancy}
\newacronym{dram}{DRAM}{Dynamic Random Access Memory}
\newacronym{dsd}{DSD}{Digital Synchronous Detection}
\newacronym{dsp}{DSP}{Digital Signal Processor}
\newacronym{dt}{DigiTime}{}
\newacronym{dvfs}{DVFS}{Dynamic Voltage and Frequency Scaling}
\newacronym{eda}{EDA}{Electronic Design Automation}
\newacronym{fdc}{FDC}{Frequency-to-Digital Converter}
\newacronym{fifo}{FIFO}{First In First Out}
\newacronym{fpga}{FPGA}{Field Programmable Gate Array}
\newacronym{gds}{GDS}{Global Data Share}
\newacronym{gnulgpl}{GNU LGPL}{GNU Lesser General Public Licence} 
\newacronym{gpgpu}{GPGPU}{General Purpose Graphics Processing Unit}
\newacronym{gpr}{GPR}{General Purpose Register}
\newacronym{gpu}{GPU}{Graphics Processing Unit}
\newacronym{gro}{GRO}{Gated Ring Oscillator}
\newacronym{io}{IO}{Input-Output}
\newacronym{hamsoc}{HAMSoC}{Hierarchical Agent Monitoring System-on-Chip}
\newacronym{hdl}{HDL}{Hardware Description Language}
\newacronym{hmp}{HMP}{Heterogeneous Multi-Processor}
\newacronym{ic}{IC}{Integrated Circuit}
\newacronym{icap}{ICAP}{Internal Configuration Access Port}
\newacronym[longplural={Intellectual Properties}]{ip}{IP}{Intellectual Property}
\newacronym{isa}{ISA}{Instruction Set Architecture}
\newacronym{lds}{LDS}{Local Data Share}
\newacronym{lru}{LRU}{Least Recently Used}
\newacronym{lsb}{LSB}{Least-Significant Bit}
\newacronym{lsu}{LSU}{Load Store Unit}
\newacronym{lut}{LUT}{Look Up Table}
\newacronym{mash}{MASH}{Multi-Stage Noise-Shaping}
\newacronym{mems}{MEMS}{Micro-Electro-Mechanical Systems}
\newacronym{miaow}{MIAOW}{Many-core Integrated Accelerator Of deepwater/Wisconsin}
\newacronym{mosfet}{MOSFET}{Metal Oxide Semiconductor Field Effect Transistor}
\newacronym{mpsoc}{MPSoC}{Multi-Processor System-on-Chip}
\newacronym{mshr}{MSHR}{Miss Status Holding/Handling Register}
\newacronym{noc}{NoC}{Network-on-Chip}
\newacronym{opencl}{OpenCL}{Open Computing Language}
\newacronym{ocn}{OCN}{On-Chip Network}
\newacronym{pcb}{PCB}{Printed Circuit Board}
\newacronym{pcie}{PCIe}{Peripheral Component Interconnect Express}
\newacronym{pl}{PL}{Programmable Logic}
\newacronym{pli}{PLI}{Verilog Programming Language Interface}
\newacronym{pll}{PLL}{Phase-Locked Loop}
\newacronym{ps}{PS}{Processing System}
\newacronym{pv}{PV}{Process Variation}
\newacronym{qoe}{QoE}{Quality of Experience}
\newacronym{qos}{QoS}{Quality of Service}
\newacronym{ram}{RAM}{Random Access Memory} 
\newacronym{risc}{RISC}{Reduced Instruction Set Computer}
\newacronym{riscv}{RISC-V}{Reduced Instruction Set Computing - V}
\newacronym{rtl}{RTL}{Register-Transfer Level}
\newacronym{sdk}{SDK}{Software Development Kit}
\newacronym{seec}{SEEC}{SElf-awarE Computing}
\newacronym{sgpr}{SGPR}{Scalar General Purpose Register}
\newacronym{si}{SI}{Southern Island}
\newacronym{simd}{SIMD}{Single Instruction Multiple Data}
\newacronym{simf}{SIMF}{Single Instruction Multiple Floating point}
\newacronym{sm}{SM}{Streaming Multiprocessor}
\newacronym{snr}{SNR}{Signal to Noise Ratio}
\newacronym[plural=SoCs,firstplural=Systems on Chip (SoCs)]{soc}{SoC}{System-on-Chip}
\newacronym{spared}{SPARED}{Self-aware PArtial Reconfiguration architecture for Edge Devices}
\newacronym{spice}{SPICE}{Simulation Program With Integrated Circuit Emphasis}
\newacronym{tad}{TAD}{Time \gls{adc}}
\newacronym[plural=TDCs,longplural={Time-to-Digital Converters}]{tdc}{TDC}{Time-to-Digital Converter}
\newacronym{tq}{TQ}{Time-Quantizer}
\newacronym{uart}{UART}{Universal Asynchronous Receiver/Transmitter}
\newacronym{vcdu}{VCDU}{Voltage Controlled Delay Unit}
\newacronym{vco}{VCO}{Voltage Controlled Oscillator}
\newacronym{vga}{VGA}{Video Graphics Array}
\newacronym{vhdl}{VHDL}{Very High Speed Integrated Circuit Hardware Description Language}
\newacronym{vlsi}{VLSI}{Very Large Scale Integration}
\newacronym{vgpr}{VGPR}{Vector General Purpose Register}
\newacronym{xilffs}{XILFFS}{Generic Fat File System Library}

\newacronym{amd}{AMD}{Advanced Micro Devices}
\newacronym{beol}{BEOL}{Back End Of Line}
\newacronym{cad}{CAD}{Computer-Aided Design}
\newacronym{eu}{EU}{European Union}
\newacronym{fdd}{FDD}{fault detection and diagnostic}
\newacronym{fefet}{FeFET}{Ferroelectric Field Effect Transistor}
\newacronym{feline}{FeLINe}{FeFET Logic IN mEmory}
\newacronym[plural=FOMs,longplural={Figures of Merit}]{fom}{FoM}{Figure of Merit}
\newacronym{hipeac}{HiPEAC}{High Performance and Embedded Architecture and Compilation}
\newacronym{hp}{HP}{Hewlett Packard}
\newacronym{hqp}{HQP}{Highly Qualified People}
\newacronym{hvac}{HVAC}{Heating, Ventilation and Air Conditioning}
\newacronym{ibm}{IBM}{International Business Machines corporation}
\newacronym{ict}{ICT}{Institute for Computer Technology}
\newacronym{iot}{IoT}{Internet of Things}
\newacronym{nda}{NDA}{Non-Disclosure Agreement}
\newacronym{nvp}{NVP}{Non-Volatile Processor}
\newacronym{oecd}{OECD}{Organization for Economic Cooperation and Development}
\newacronym{rd}{R\&D}{Research and Development}
\newacronym{soa}{SoA}{State-of-the-Art}
\newacronym{tsmc}{TSMC}{Taiwan Semiconductor Manufacturing Company}
\newacronym{tvlsi}{TVLSI}{Transactions on Very Large Scale Integration}

\newacronym{cas}{CAS}{Compare-and-Swap}
\newacronym{six}{SIXOR}{Single-cycle In-Memristor XOR}
\newacronym{nvm}{NVM}{Non-Volatile Memory}

\usepackage{subfigure}
\usepackage{stackengine}

\fancypagestyle{firstpage}
{
    \fancyhead[R]{}
    \fancyhead[C]{\textbf{\footnotesize This Article has been Accepted for Publication in ACM Journal on Emerging Technologies in Computing Systems}}
}
\begin{document}

\title{Sorting in Memristive Memory}


\author{Mohsen Riahi Alam}
\affiliation{%
  \institution{University of Louisiana at Lafayette}
  \state{Louisiana}
  \country{USA}}

\author{M. Hassan Najafi}
\affiliation{%
  \institution{University of Louisiana at Lafayette}
  \state{Louisiana}
  \country{USA}}

\author{Nima TaheriNejad}
\affiliation{%
  \institution{TU Wien}
  \city{Vienna}
  \country{Austria}}

\renewcommand{\shortauthors}{This Article has been Accepted for Publication in ACM Journal on Emerging Technologies in Computing Systems}

\thanks{This work was supported in part by the Louisiana Board of Regents Support Fund no. LEQSF(2020-23)-RDA-
26, National Science Foundation grant \#2019511, and generous gift from Cisco}


\begin{abstract}
Sorting data is needed in many application domains. Traditionally, the data is read from memory and sent to a general-purpose processor or application-specific hardware for sorting. The sorted data is then written back to the memory. Reading/writing data from/to memory and transferring data between memory and processing unit incur significant latency and energy overhead.  In this work, we develop the first architectures for in-memory sorting of data to the best of our knowledge. We propose two architectures. The first architecture is applicable to the conventional format of representing data, i.e., weighted binary radix. The second architecture is proposed for developing unary processing systems, where data is encoded as uniform unary bit-streams. As we present, each of the two architectures has different advantages and disadvantages, making one or the other more suitable for a specific application. However, the common property of both is a significant reduction in the processing time compared to prior sorting designs. Our evaluations show on average 37$\times$ and 138$\times$ energy reduction for binary and unary designs, respectively, compared to conventional CMOS off-memory sorting systems in a 45nm technology. We designed a $3\times3$ and a $5\times5$ Median filter using the proposed sorting solutions, which we used for processing $64\times64$ pixel images. Our results show a reduction of 14$\times$ and 634$\times$ in energy and latency, respectively, with the proposed binary, and 5.6$\times$ and 152$\times10^3$ in energy and latency with the proposed unary approach compared to those of the off-memory binary and unary designs for the 3$\times$3 Median filtering system.
\end{abstract}


\maketitle

\glsresetall

\thispagestyle{firstpage}
\section{Introduction}
\label{sec_intro}
Sorting is a fundamental operation in computer science, used in databases~\cite{ApplicationDatabase,ApplicationDatabase2}, scientific computing~\cite{ApplicationScientific}, scheduling~\cite{ApplicationScheduling}, artificial intelligence and robotics~\cite{ApplicationRobotic}, image~\cite{Peng_TVLSI14}, video~\cite{ApplicationVideo}, and signal processing~\cite{ApplicationSignal}. The latency and energy consumptions of the sorting algorithm directly affect the efficiency of these systems.
A sizeable body of research has focused on harnessing the computational power of many-core \gls{cpu}- and \gls{gpu}-based  systems for efficient sorting~\cite{Sorting_GPU1,Sorting_GPU2,Sorting_GPU3}. For high-performance applications, sorting is implemented in hardware using either \glspl{asic} or \glspl{fpga}~\cite{Review2Ref3,Sorting_ASIC,Sorting_FPGA1}. The parallel nature of hardware-based solutions allows them to outperform software-based solutions executed on \glspl{cpu}/\glspl{gpu}.

\begin{figure}[!t] 
\centering
\includegraphics[clip,trim={0cm 0cm 12cm 26cm},width=4.5in]{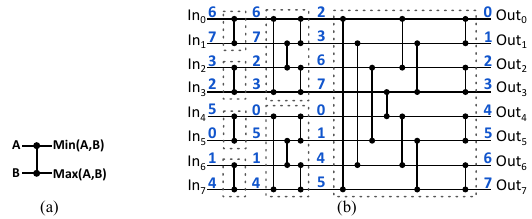}
\caption{(a) Schematic symbols of a \gls{cas} block (b) \gls{cas} network for an 8-input bitonic sorting.}
\label{Bitonic8}
\end{figure}

\begin{figure}[!t] 
\centering
\small (a)\includegraphics[width=1.95in]{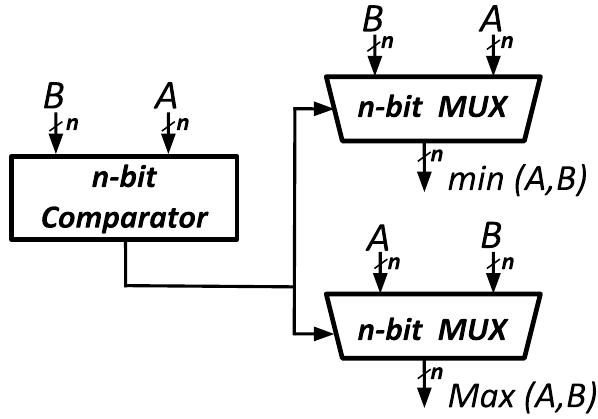}~~~~~
\small \color{white}...\color{black}~~~~~(b)\includegraphics[clip,trim={0.0cm -.5cm 0cm 0.0cm } , width=2.2in]{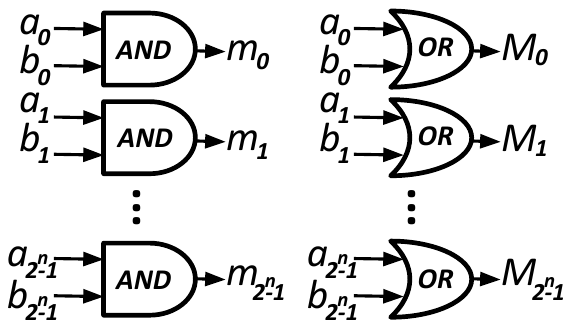}
\caption{Logic design of a \gls{cas} block: (a)~conventional binary design processing data (b)~parallel unary design processing unary bit-streams.}
\label{Fig:CAS}
\end{figure}

The usual approach for hardware-based sorting is to wire up a network of \gls{cas} units in a configuration called a Batcher (or bitonic) network~\cite{SortingBatcher}. Batcher networks provide low-latency solutions for hardware-based sorting~\cite{SortingPaper1,Review2Ref1}. Each \gls{cas} block compares two input values and, if required, swaps the values at the output. Fig.~\ref{Bitonic8}(a) shows the schematic symbol of a \gls{cas} block. Fig.~\ref{Bitonic8}(b) shows the \gls{cas} network for an 8-input bitonic sorting network, made up of 24 \gls{cas} blocks. Batcher sorting networks are fundamentally different from software algorithms for sorting (such as the quicksort, merge sort, and the bubble sort), since the order of comparisons is fixed in advance. That is, in contrast to software algorithms, the order is data-dependent ~\cite{Sorting-TVLSI-2018}. The implementation cost of a batcher network is a direct function of the number of \gls{cas} blocks and the cost of each block. A \gls{cas} block is conventionally designed based on the weighted binary radix representation of data. The \gls{cas} design consists of an $n$-bit comparator and two $n$-bit multiplexers, where $n$ is the data-width of the input data\footnote{We note that later on, we take the precision of binary to unary conversion (for unary sorting solution) to be also equal to $n$, the data-width.}.  Fig.~\ref{Fig:CAS}(a) shows the conventional\ignore{ binary radix}design of a \gls{cas} unit. In the conventional binary design, increasing the data-width increases the complexity of the design.

All these prior sorting designs were developed based on the Von-Neumann architecture, separating the memory unit where the data is stored and the processing unit. where the data is processed (i.e., sorted). 
A significant portion of the total processing time and the total energy consumption is wasted on 1) reading the data from memory, 2) transferring the data between memory and processing unit, and 3) writing the result back into the memory ~\cite{NDPMicro14}~\cite{memory_energy_wire}~\cite{Ruan_ICCAD19}~\cite{Koo_Micro17}~\cite{Tseng2015GullfossA}.
\gls{imc} or \gls{pim} is an emerging computational approach that offers the ability to both store and process data within memory cells~\cite{Zidan2018,Taherinejad2015,Radakovits2019,Hamdioui2017DATE,xie2017scouting, Taherinejad2016,Scrimp_Gupta20,Riahi_DesignTest21}. This technique eliminates the high overhead of transferring data between memory and processing unit, improving the performance and reducing the energy consumption by processing data in memory. 
Memristive storage is a \gls{nvm} with high storage density and \gls{imc} capability. This emerging technology is one of the most promising candidates for the next generation of storage systems. The \gls{imc} capability of \gls{nvm} devices allows accelerating sorting by avoiding the overheads of transferring the data between memory and processing unit. New sorting approaches based on \gls{nvm} technology are on the table to increase the efficiency of the hardware-based sorters. 
Some previous studies worked on optimizing sorting algorithms for \gls{nvm} and presented \gls{nvm}-friendly sorting algorithms~\cite{Khernache17}~\cite{chunvmsorting}. Prasad~et~al.~\cite{Prasad_HPCA21} proposed RIME which provides an API library for sorting algorithms using a bit-level search operation within the memory. They use some additional CMOS circuitry including a sensing circuit to compute the minimum and maximum values by performing XOR operation on the memory peripheral circuits. Some previous studies focused on near storage/memory computing techniques for providing efficient sorting algorithms. Li~et~al.~\cite{Li_GLSVLSI20} proposed IMC-Sort, an in-memory parallel sorting architecture using the hybrid memory cube. Their architecture incorporates a custom parallel sorting unit to accelerate the sort workloads in DRAM based on 3D stacking technology. Pugsey~et.~al.~\cite{ICCD2015} suggested 3D-stacked near-data processing for sorting data in DRAM. Processing units and memory are integrated with 3D stacking technology using through-silicon vias. Salamat~et~al.~\cite{salamat2021nascent} proposed a near-storage accelerator for databases sort based on the bitonic sort. Their accelerator utilizes an NVMe flash drive with an onboard FPGA chip. The authors in~\cite{Qiao21} propose FANS; an FPGA accelerated near-storage sorting system. Their system is able to sort hundreds of gigabytes of data on a single Samsung SmartSSD. The authors in~\cite{bonsai2020} introduced Bonsai, an adaptive FPGA-based near-memory sorting solution. Their design considers the off-chip memory bandwidth and on-chip resources to optimize sorting time. Casper and Olukotun ~\cite{ICCD2015} presented three hardware accelerator designs to perform near-memory database operations including sort. Evaluation results by implementing their designs on FPGA showed close to ideal utilization of available memory bandwidth. These prior works sort the data near-memory or in-memory within peripheral circuitry. None of them perform sorting in memory within the memory array. For a detailed classification of different near- and in-memory computing methods, the readers are referred to~\cite{Class20}.

In this paper, we take advantage of \gls{imc} to implement sorting units on memristive memory arrays. To the best of our knowledge, this work introduces the first \textit{in-array} architectures for high-performance and energy-efficient sorting of data completely in memory (CIM-A using the~\cite{Class20} terminology). Our work is different from the aforementioned prior works in the sense that all computation results are produced within the memory array. We then go further to show how we can benefit from the concept of Unary Computing~\cite{Poppelbaum198747,Sorting-TVLSI-2018} to improve the sorting hardware further for particular applications. 
We propose two different architectures. The first architecture, "Binary Sorting," is  based on the conventional weighted binary representation and is applicable to  conventional systems that store the data in memory in the binary format. The second architecture, "Unary Sorting," is based on the non-weighted unary representation.
For each of these designs, we first discuss the basic operation of sorting two $n$-bit data (i.e., a \gls{cas} block). We then elaborate on the design of complete sorting networks, which are made up of the proposed in-memory \gls{cas} units. We showcase the role and importance of the achieved gains in the context of a median filter used in image processing applications. Our experiments demonstrate a reduction of 14$\times$ and 634$\times$ in energy and latency for the proposed binary, and 5.6$\times$ and 152$\times10^3$ in energy and latency for the proposed unary approach compared to those of the off-memory binary and unary designs when implementing a 3$\times$3 Median filtering system.

The rest of this paper is structured as follows. \Cref{sec_background} provides a brief overview of memristive \gls{imc} and the unary processing technique used in this work. 
\Cref{sec_proposed} and \Cref{sec_prop_unary} present the proposed in-memory Binary and Unary Sorting designs. \Cref{sec_comp} compares the performance of the proposed designs with the conventional off-memory CMOS-based designs and applies the proposed architectures to an important application of sorting, i.e., median filtering. Finally, conclusions are drawn in Section~\ref{sec_conclusions}.

\section{Background}
\label{sec_background}
\subsection{Memristive IMC}

\begin{figure}[t]
\centering
\includegraphics[clip,trim={0.5cm 0.0cm 10cm 22cm},width=4.0in]{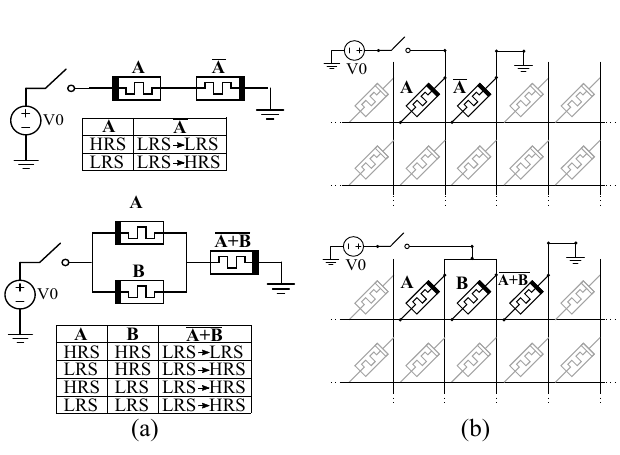}
\vspace{-0.5em}
\caption{{(a) \texttt{NOT} and \texttt{NOR} logical operations in MAGIC and their truth tables. \gls{lrs} and \gls{hrs} represent logical ‘1’ and logical ‘0’, respectively.  (b) Crossbar implementation of NOT and NOR logical operations.}}
\label{fig:magic-not-nor}
\end{figure}

One of the promising technologies for \gls{imc} is memristive technology. Among various memristive-based \gls{imc} methods, stateful logic such as \gls{imply}~\cite{borghetti2010nature},  \gls{magic}~\cite{MAGIC_2014}, FELIX~\cite{Felix_imani}, and \gls{six}~\cite{Taherinejad2021tvlsi} are of the most efficient solutions. In stateful logic, the input and output are both presented as the state of input and output memristors. Hence, no access to the world outside the array (e.g., read or write) is necessary for stateful logic operations. In this work, we use \gls{magic} \texttt{NOR} operation, which  can be used to implement any Boolean logic. \gls{magic} considers two states of memristors: \gls{lrs} as logical ‘1’ and \gls{hrs} as logical ‘0’. Fig.~\ref{fig:magic-not-nor}(a) shows how \texttt{NOR} and \texttt{NOT} logical operations can be implemented in MAGIC~\cite{MAGIC_2014}, where the  memristors connected to the ground are output memristors~\cite{MAGIC_2014}. Before starting the execution of an operation, the output memristors are first  initialized to \gls{lrs}. By applying a specific voltage ($V_0$) to the negative terminal of the input memristors, the output memristors may experience a state change from \gls{lrs} to \gls{hrs}, depending  on the states of the inputs~\cite{MAGIC_2014}. The truth tables embedded  in  Fig.~\ref{fig:magic-not-nor}(a) show  all  possible  cases of the input memristors' states and switching of the output memristors. Fig.~\ref{fig:magic-not-nor}(b) shows how \gls{magic} \texttt{NOT} and \texttt{NOR} can be realized in a crossbar memory. These operations can be natively executed within memory with a high degree of parallelism. Thus, parallel architectures such as sorting networks can benefit greatly from such \gls{imc} logic operations.

\subsection{In memory Comparator}
Comparison is an essential operation in  implementing sorting functions. Comparing memory content has been always challenging in computing systems. Content-addressable memory (CAM) ~\cite{CAM2006} uses a dedicated equality comparator circuit to return the location of the matching data. CAMs help searching architectures and can be applied to packet forwarding in network routers. Authors in ~\cite{Prasad_HPCA21} propose a method for finding the minimum and maximum values within a set of numbers in memory. They employ bitwise column search to design a bit-serial algorithm for finding the minimum and maximum value. The sorting mechanisms proposed in~\cite{Li_GLSVLSI20} and ~\cite{ICCD2015} are based on a bitonic sorting network and include some comparison units. The comparison units in these works are implemented at the logic layer, which is integrated with 3D stacking DRAM memory banks using through-silicon vias technology. The authors in ~\cite{EqualityComparator1} and \cite{EqualityComparator2} further propose two in-memory equality comparators for SRAM memories.

Authors in ~\cite{ED2020} developed a multivalued 1T1R memristor method for in-memory computing. They exploit the multivalued resistance for performing a 1-bit in-memory comparison and then expand the design to a 4-bit magnitude comparator. Angizi et~al propose an in-memory magnitude comparator in~\cite{Angizi_DAC19}. Their design uses in-memory \texttt{XOR} operations to perform a bit-wise comparison between corresponding bits of two data beginning from the most significant bit towards the least significant bit. However, the comparison process involves reading the output of the \texttt{XOR} operations and the data from memory by the control unit.
Therefore, its latency (i.e., number of processing cycles) is non-deterministic and depends on the data being compared. In Section~\ref{Sec:Basic_Sorting_binary}, we propose an in-memory magnitude comparator with deterministic latency and no memory read operations (a stateful comparator). Our in-memory comparator does not also need multivalued memristors.

\subsection{Unary Sorting}

Unary (or burst) processing~\cite{Poppelbaum198747,PoppelbaumUnary2} is an alternative computing paradigm to conventional binary offering simple and noise-tolerant solutions for complex arithmetic functions~\cite{Smith_ISCA_2018,ISCAS20_Jalilvand_Fuzzy,RoutingMagic_FPGA_2018,Sorting-TVLSI-2018,Faraji_Hybrid_TC,RaceLogic,Temporal_ASPLOS20,Najafi_Pulse_ISCAS20,UGEMM2020}. The paradigm borrows the concept of averaging from stochastic computing~\cite{Gaines1969,Armin_Survey_2018}, but is deterministic and accurate.
In unary processing, unlike weighted binary radix, all digits are weighted equally.   Numbers are encoded uniformly by a sequence of one value (e.g., 1) followed by a sequence of the other value (e.g., 0) in a stream of 1's and 0's--\ignore{The streams of this format are }called a \textit{unary bit-stream}. The value of a unary bit-stream is determined by the frequency of the appearance of 1’s in the bit-stream.\ignore{The real number $i/N$ can be represented using a total of $N$ bits, in which $i$ bits are ones and $N-i$ bits are zeros. }For example, 11000000 is a unary bit-stream representing 2/8 or 1/4.

Unary computing was first used in~\cite{Sorting-TVLSI-2018,Najafi_Sorting_ICCD2017} for the simple and low-cost implementation of sorting network circuits. Zhang~et~al.~\cite{Zhang_DATE2020} developed an SC-based neural network accelerator by employing a bit-stream-based bitonic sorting network for simultaneously implementing the accumulation and activation functions. With unary bit-streams and also when using correlated stochastic bit-streams~\cite{alaghi_Correlation}, minimum and maximum functions (the main operations in a \gls{cas} block) can be implemented using simple standard \texttt{AND} and \texttt{OR} gates. In a serial manner, one \texttt{AND} and one \texttt{OR} gate implements a \gls{cas} block by processing one bit of the two bit-streams at each cycle. Hence, a total of $2^n$ processing cycles is needed to process two $2^n$-bit bit-streams (equivalent to two $n$-bit binary data since we chose the precision of binary to unary conversion to be equal to the data-width, that is, equal to $n$). More than 90\% saving in the hardware cost is reported for a 256-input serial unary sorting circuit at the cost of processing time~\cite{Sorting-TVLSI-2018}. Alternatively, the bit-streams can be processed in one cycle by replicating the logic gates and performing the logical operations in parallel. Fig~\ref{Fig:CAS}(b) shows the parallel unary design of a CAS block. $2^n$ pairs of \texttt{AND} and \texttt{OR} gates sort two $2^n$-bit bit-streams. 

\section{Proposed In-Memory Binary Sorting}
\label{sec_proposed}
In this section, we present our proposed method for in-memory sorting of binary radix data. First, we discuss the implementation of a basic sorting unit and then generalize the architecture to complete sort systems.

\begin{figure}[t]
\centering
\small (a)~~\includegraphics[clip,trim={0.cm 0.1cm 16cm 25.0cm}, width=1.7in]{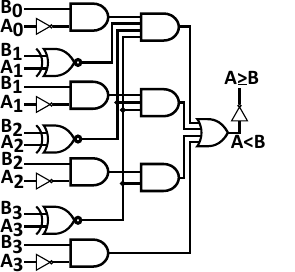}
(b)~~\includegraphics[clip,trim={0.cm 0cm 18.3cm 25.9cm}, width=1.15in]{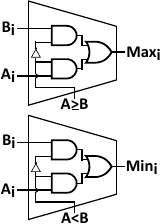}

\caption{Generic logic design of (a) a 4-bit binary magnitude comparator and (b) a multi-bit binary 2-to-1 multiplexer for Max/Min selection.}
\label{Fig:4-bit-Comparator-MUX-Binary}
\end{figure}

\begin{figure} [!t]
\centering
\small (a)~~~\subfigure{\includegraphics[clip,trim={0.0cm 0.05cm 14.3cm 25.1cm}, width=2.5in]{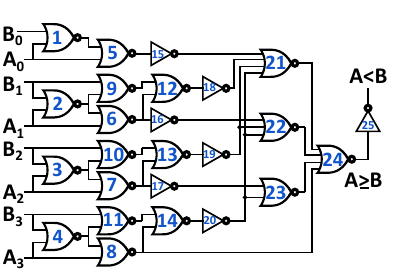}
}
\small (b)~~~\subfigure{\includegraphics[clip,trim={0.0cm 0cm 6cm 25.1cm}, width=5.5in]{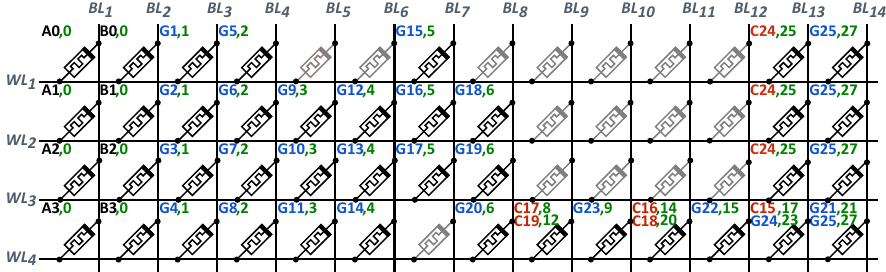}
}
\caption{(a) \texttt{NOR}-based logic design of a 4-bit binary comparator. (b) MAGIC-based 4-bit binary in-memory comparator. \blue{$G_i$}~memristor holds the output of the \blue{$i$}-th gate. \red{$C_i$}~memristor copies the state of \blue{$G_i$}~memristor. The second number shown on each memristor (e.g., \textbf{\green{2}} in \blue{G5},\textbf{\green{2}}) determines the processing cycle in which the memristor operates. (WL = Word Line, BL = Bit Line)}
\label{MAGIC-4-bit-Comp-Binary}
\end{figure}

\begin{figure*} [!t]
\centering
\small (a)~~~\subfigure{
\includegraphics[clip,trim={0.cm -0.1cm 15.5cm 26.8cm}, width=2.1in]{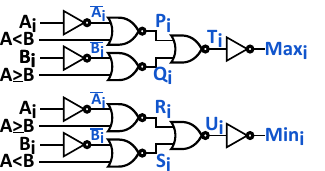}
\label{Fig:NOR-4-bit-MUX-Binary}
}
\small (b)~~~\subfigure{
\includegraphics[clip,trim={0.cm 0cm 8.8cm 25.0cm}, width=4.8in]{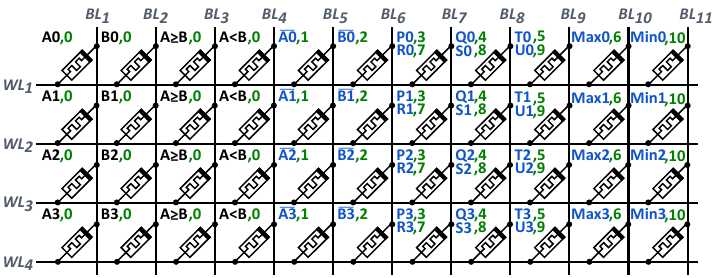}
\label{Fig:MAGIC-4-bit-MUX}
}
\caption{(a) \texttt{NOR}-based logic design of a multi-bit binary 2-to-1 multiplexer and (b) in-memory MAGIC-based 4-bit binary multiplexer for Max/Min selection. The second number shown on each memristor (e.g., \textbf{\green{3}} in \blue{P0}\textbf{\green{,3}}) determines the processing cycle in which the memristor operates. (WL = Word Line, BL = Bit Line)}
\label{fig:4-bit-MUX-all}
\end{figure*}

\subsection{Basic Binary Sorting Unit}
\label{Sec:Basic_Sorting_binary}

A basic binary sorting unit (\gls{cas} unit) requires one comparator and two multiplexers. Implementing an $n$-bit comparator by using basic logic gates requires ($11n-2$) \texttt{NOR} and ($7n-2$) \texttt{NOT} logic gates.
Figs.~\ref{Fig:4-bit-Comparator-MUX-Binary}(a) and \ref{MAGIC-4-bit-Comp-Binary}(a) show the generic logic and the \texttt{NOR}-based logic design of a 4-bit binary comparator.
Fig.~\ref{MAGIC-4-bit-Comp-Binary}(b) shows our proposed  in-memory implementation using \gls{magic}. 
As shown, implementing this comparator using MAGIC \texttt{NOR} and \texttt{NOT} operations requires a crossbar with $4\times14$ memory cells.
The input data (i.e., A and B) in binary format is stored in two different columns (BL$_1$ and BL$_2$), each column containing $n$ memristors, where $n$ is the size of the data being compared (in this example, $n=4$).  The computation includes \texttt{NOR}, \texttt{NOT}, and copy operations. Each $G_{i,j}$ memristor in Fig.~\ref{MAGIC-4-bit-Comp-Binary}(b) shows participation in a logical gate (operation) $i$ in Fig.~\ref{MAGIC-4-bit-Comp-Binary}(a). $C_{i,j}$s on the other hand, show the copy operation $i$, in which the state of $G_i$ memristor is duplicated. The index $j$ marks the cycle number in which an operation is performed. In some cases, a memristor participates in two operations. For example, the memristor at the right-bottom end of Fig.~\ref{MAGIC-4-bit-Comp-Binary}(b) (WL$_4$, BL$_{14}$) is once used at cycle 21 in gate (operation) 21 and once at cycle 27 in gate (operation) 25.

To execute these operations, in each clock cycle, the memristor controller applies the proper voltage to crossbar columns and rows to execute some \texttt{NOR} or \texttt{NOT} operations concurrently. All memristors with the same cycle number produce their output value at the same time. When possible, we reuse memristors to avoid increasing area, i.e., the number of used memristors. The memristors that are being re-used as an output must be initialized to \gls{lrs} in an additional clock cycle before reusing. The comparison result (i.e., the output of gate~24) is ready at cycle~23 on (WL$_4$, BL$_{13}$). At this time, some copies from the comparison result and its complement must be made. These will be used as the select inputs of the maximum and minimum multiplexers shown in  Fig.~\ref{Fig:4-bit-Comparator-MUX-Binary}(b) and Fig.~\ref{fig:4-bit-MUX-all}. To this end, first, we make three copies of the output of gate~24 on three memristors in the same column (BL$_{13}$) and then invert these memristors on another column (BL$_{14}$) to make the required complements. This leads to a total processing time of  27 cycles plus one initialization cycle. 

\ignore{
\begin{figure}[t]
\centering
\subfigure[]{
\includegraphics[clip,trim={0.cm 0cm 15.5cm 26.2cm}, width=1.5in]{figures/bin_comp6.pdf}}\hspace{-0.3em}\subfigure[]{
\includegraphics[clip,trim={0.cm 0cm 15.5cm 27.cm}, width=1.8in]{figures/bin_comp4.pdf}
\label{Fig:NOR-4-bit-MUX-Binary}
}
\vspace{-1.5em}
\caption{\texttt{NOR}-based logic design of a multi-bit binary 2-to-1 multiplexer for min/max selection.}
\label{Fig:4-bit-MUX-Binary}
\end{figure}
}

After the comparison process, we need values of only four columns (BL$_1$ and BL$_2$ for the two input data, and BL$_{13}$ and BL$_{14}$ for the two comparison results) to implement the multiplexer part of the sorting unit. Hence, we could reuse the rest of the memristors. Fig.~\ref{Fig:4-bit-Comparator-MUX-Binary}(b) and Fig.~\ref{fig:4-bit-MUX-all} show the generic logic  and the \texttt{NOR}-based logic circuit for a multi-bit 2-to-1 multiplexer. Fig.~\ref{Fig:MAGIC-4-bit-MUX} shows our MAGIC-based in-memory design for the two 4-bit multiplexers the sorting circuit requires to select the maximum and minimum data. 
In implementing the multiplexers, we re-use the memory cells of the comparison step. To this end, we initialize the columns used in the comparison step (BL$_3$ to BL$_{12}$) to \gls{lrs} in one clock cycle. The input data is inverted in two clock cycles, cycles 1 and 2 (on BL$_4$ and BL$_5$) shown in Fig.~\ref{Fig:MAGIC-4-bit-MUX}. The first multiplexer produces the maximum value on BL$_{10}$ in cycles 3 to 6. The minimum value is produced on BL$_{11}$ by the second multiplexer through cycles 7 to 10. Since three columns used by the first multiplexer (i.e., P, Q, T) are being re-used by the second multiplexer, an additional cycle is considered for the initialization of these columns before execution of the second multiplex operation. The execution of the multiplexers, therefore, takes two initialization and 10 operation cycles. Hence, execution of the proposed in-memory basic binary sorting takes a total of 39 processing cycles plus one initialization cycle.

We extend the proposed design from sorting of 4-bit data to higher data-widths, namely 8-, 16-, 32-, and in general $n$-bit data. We verified the correct functionality of the proposed design by high-level simulation and measured the energy and delay numbers by circuit-level simulation using Cadence Virtuoso. 
Table~\ref{Table:BasicBinary} reports the required resources, the number of cycles, and energy consumption. Further details on circuit-level simulations and the parameter values used in estimating the energy numbers will be discussed in Section \ref{Sec:CirSimulation}. We see that the area, the latency, and the energy consumption of the proposed basic binary sorting design increase linearly by increasing the data-width.

\begin{table}[!t]
\centering
\caption{{The Required Resources and Number of Processing Cycles for the Proposed Basic Binary Sorting Unit}}
\label{Table:BasicBinary}
\setlength{\tabcolsep}{3pt}
\renewcommand{\arraystretch}{1.}
\resizebox{16cm}{!}
{
\begin{tabular}{|c|c|c|c|c|c|c|c|c|}
\hline
\multirow{2}{*}{Data-Width} & Required & \# of Initialized & \# of Reused & \# of Logical & \# of Copies & Reused init  & Total \# of & Energy \\
                           & \multicolumn{1}{c|}{Dimension} & \multicolumn{1}{c|}{Memristors} & \multicolumn{1}{c|}{Memristors} & \multicolumn{1}{c|}{Operation Cycles} & \multicolumn{1}{c|}{{[}Copy Cycles{]}} & \multicolumn{1}{c|}{Cycles} & 
                           \multicolumn{1}{c|}{Cycles} &
                           \multicolumn{1}{c|}{($pJ$)} \\ \hline
4 & 4$\times$14 & 40 & 44 & 21 & 6 (12) & 6 & 39 & 199.4\\ \hline
8 & 8$\times$22 & 88 & 88 & 25 & 14 (28) & 10 & 63 & 417\\ \hline
16 & 16$\times$38 & 184 & 176 & 33 & 30 (60) & 18 & 111 & 845\\ \hline
32 & 32$\times$70 & 376 & 352 & 49 & 62 (124) & 34 & 207 & 1728\\ \hline
$n$  & $n$ $\times$ (8+2$n$-2) & $12n-8$ & $11n$ & 18+($n$-1) & 2$n$-2 {[}2(2$n$-2){]} & $n+2$ & 6$n+15$ & -\\ \hline
\end{tabular}
}
\end{table}

\subsection{Complete Binary Sort System}
\label{Sec:Binary-Complete-sort}

A complete sort network  is made of basic sorting units (i.e., \gls{cas} blocks). In bitonic sorting, the network recursively merges two sets of size $N/2$ to make a sorted set of size
$N$~\cite{SortingPaper2}. Fig.~\ref{Bitonic8} shows the \gls{cas} network for an 8-input bitonic sorting. As it can be seen, the network is made of 24 \gls{cas} units. In general, an $N$-input bitonic sorting network requires 
\begin{equation}
  {U_{CAS}} = N \times log_2(N) \times (log_2(N)+1)/4
\end{equation}
\gls{cas} units. These CAS units can be split into 
\begin{equation}
   S = log_2(N)\times(log_2(N)+1)/2 
\end{equation}
 steps
 (also known as \textit{stages}), each with $N$/2 CAS units that can operate in parallel~\cite{SortingPaper1}.

\begin{table*}[t]
\centering
\caption{{Number of Processing Cycles, Size of Crossbar Memory, and Energy Consumption ($nJ$) to Implement Different\\ Bitonic Sorting Networks (DW = Data-Width, BL = Bit-Stream Length)}}
\label{Table:SortingNetwork-Lateny-Dimension}
\setlength{\tabcolsep}{5pt}
\renewcommand{\arraystretch}{1.2}
\resizebox{16cm}{!}
{
\begin{tabular}{|c|c|c|c|c|c|c|c|c|c|c|c|c|}
\hline
\multirow{2}{*}{\begin{tabular}[c]{@{}c@{}}Network\\ Size\end{tabular}} & \multicolumn{3}{c|}{\textbf{Binary Sorting} DW = 4} & \multicolumn{3}{c|}{DW = 8} & \multicolumn{3}{c|}{DW = 16} & \multicolumn{3}{c|}{DW = 32} \\ \cline{2-13} 
 & Cycles & Size & Energy & Cycles & Size & Energy & Cycles & Size & Energy & Cycles & Size & Energy \\ \hline
4 & 128 & 4$\times$28 & 1.2 & 200 & 8$\times$44 & 2.5 & 344 & 16$\times$76 & 5.1 & 632 & 32$\times$140 & 10 \\ \hline
8 & 280 & 4$\times$56 & 4.7 & 424 & 8$\times$88 & 10 & 712 & 16$\times$152 & 20 & 1288 & 32$\times$280 & 41 \\ \hline
16 & 544 & 4$\times$112 & 15 & 784 & 8$\times$176 & 33 & 1264 & 16$\times$304 & 68 & 2224 & 32$\times$560 & 138 \\ \hline
32 & 1048 & 4$\times$224 & 47 & 1408 & 8$\times$352 & 100 & 2128 & 16$\times$608 & 205 & 3568 & 32$\times$1120 & 415 \\ 
\hline\hline
\multirow{2}{*}{\begin{tabular}[c]{@{}c@{}}Network\\ Size\end{tabular}} & \multicolumn{3}{c|}{\textbf{Unary Sorting} BL = 16} & \multicolumn{3}{c|}{BL = 64 {(DW = 6)}} & \multicolumn{3}{c|}{BL = 256 {(DW = 8)}} & \multicolumn{3}{c|}{BL = 1024 {(DW = 10)}} \\ \cline{2-13} 
 & Cycles & Size & Energy & Cycles & Size & Energy & Cycles & Size & Energy & Cycles & Size & Energy \\ \hline
4 & 26 & 16$\times$10 & 1.37 & 26 & 64$\times$10 & 5.4 & 26 & 256$\times$10 & 21.88 & 26 & 1024$\times$10 & 87 \\ \hline
8 & 76 & 16$\times$20 & 5.4 & 76 & 64$\times$20 & 21 & 76 & 256$\times$20 & 87 & 76 & 1024$\times$20 & 350 \\ \hline
16 & 194 & 16$\times$40 & 18 & 194 & 64$\times$40 & 72 & 194 & 256$\times$40 & 291 & 194 & 1024$\times$40 & 1168 \\ \hline
32 & 538 & 16$\times$80 & 54 & 538 & 64$\times$80 & 218 & 538 & 256$\times$80 & 875 & 538 & 1024$\times$80 & 3503 \\ \hline
64 & 1406 & 16$\times$160 & 153 & 1406 & 64$\times$160 & 613 & 1406 & 256$\times$160 & 2452 & 1406 & 1024$\times$160 & 9809 \\ \hline
128 & 3624 & 16$\times$320 & 408 & 3624 & 64$\times$320 & 1635 & 3624 & 256$\times$320 & 6540 & 3624 & 1024$\times$320 & 26159 \\ \hline
256 & 9176 & 16$\times$640 & 1051 & 9176 & 64$\times$640 & 4204 & 9176 & 256$\times$640 & 16817 & 9176 & 1024$\times$640 & 67268 \\ \hline
\end{tabular}
}
\end{table*}

\begin{figure}[t]
\centering
\includegraphics[clip,trim={0.0cm 0.0cm 8cm 14cm},width=3.2in]{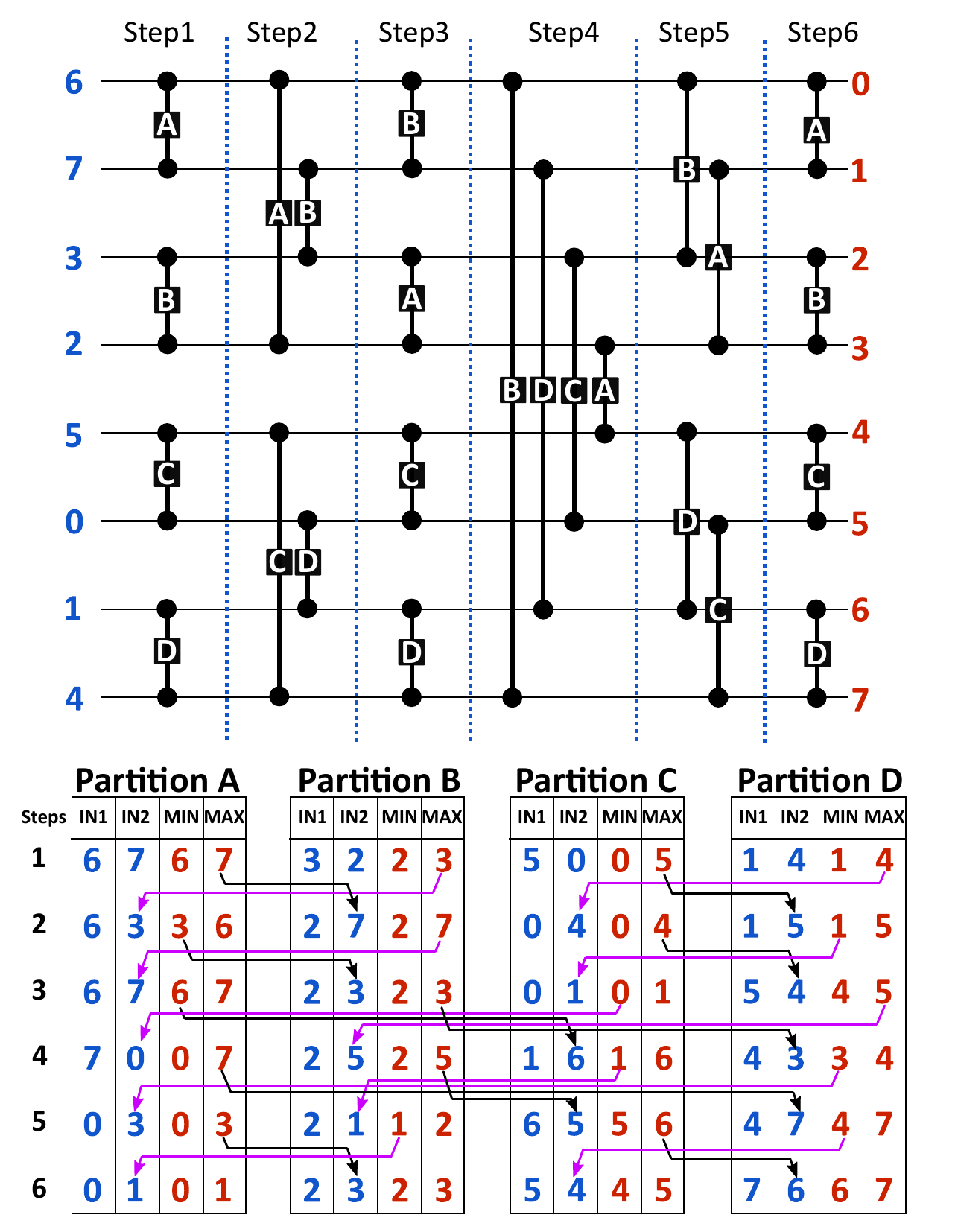}
\caption{High-level flow of 8-input Bitonic Sorting in Memory.}
\label{fig:8-inputSortingFlow}
\end{figure}

Gupta et al~\cite{Felix_imani} propose a memory partitioning method to improve the in-memory parallelism. 
In a similar fashion, we split the memory into multiple partitions to enable parallel execution of different CAS operations in each bitonic CAS stage. Fig.~\ref{fig:8-inputSortingFlow} 
shows how we implement an 8-input bitonic sorting network in memory. The memory is split into four partitions, namely partitions A, B, C, and D (each marked on a black vertical line in the bitonic network representation). The number of partitions is decided based on the number of CAS units that can run in parallel (i.e., $N/2$.). Each partition includes two out of the eight unsorted input data. The sorting process is split into six steps equal to the number of CAS groups (stages). In the first step, the two inputs in each partition are sorted using the basic sorting operation proposed in Section II-A. In the second step, each maximum number (i.e., the larger number between the two in the partition) found by the sorting operations of the first step is copied to another partition where it is needed. The bitonic network determines the destination partition. For instance, the maximum found by executing the sorting operation in partition A (i.e., the input with a value of 7 in the example of Fig.~\ref{fig:8-inputSortingFlow}) will be copied into partition B to be compared with the minimum number between the two initial data in partition B of the first step. Similarly, in each one of the next steps (i.e., steps 3 to 6), one output data from each partition is copied to another partition, and a sorting operation is executed.

In each step, the sortings in different partitions are executed in parallel. After six steps and the execution of a total of 24 (=$4\times6$) basic sorting operations, the sorted data is ready in the memory.\ignore{Table~\ref{Table:SortingNetwork-Steps} compares the number of sorting steps and the total number of basic sorting units for implementation of different bitonic sorting networks with sizes of 4, 8, 16, and general case of $N$ inputs. }Each basic sorting operation is implemented based on the in-memory basic binary sorting proposed in Section~\ref{Sec:Basic_Sorting_binary}. Table~\ref{Table:SortingNetwork-Lateny-Dimension} shows the total number of processing cycles, the required size of crossbar memory, and the energy consumption of different sizes of in-memory bitonic networks. {The total number of processing cycles,  $PC_t$, is calculated using }
\begin{equation}
    PC_t=S\times(1+PC_b)+CP,
\end{equation}
where $PC_b$ is the number of processing cycles necessary to execute a basic sorting operation, 
$CP$ is the number of copy operations, and $S$ the number of 
sorting steps.
The required size of crossbar memory ($M_t$) is found by
\begin{equation}
    M_t=n\times \frac{N}{2} \times M_b,
\end{equation}
where $M_b$ is the size of the crossbar memory required for one basic sorting unit\footnote{We remember that $n$ is the data-width and $N$ is the network size or the total number of items to be sorted.}.

\ignore{
\begin{table}[]
\begin{tabular}{|c|c|c|c|c|}
\hline
\multirow{2}{*}{Network} & \multicolumn{4}{c|}{Binary Energy (nJ)} \\ \cline{2-5} 
 & 4 & 8 & 16 & 32 \\ \hline
4 & 0.132 & 0.276 & 0.565 & 1.143 \\ \hline
8 & 0.54 & 1.108 & 2.265 & 4.58 \\ \hline
16 & 1.801 & 3.696 & 7.557 & 15.278 \\ \hline
32 & 5.407 & 11.092 & 22.679 & 45.85 \\ \hline
\multirow{2}{*}{Network} & \multicolumn{4}{c|}{Unary Energy (nJ)} \\ \cline{2-5} 
 & $2^4$ & $2^6$ & $2^8$ & $2^{10}$ \\ \hline
4 & 0.150 & 0.599 & 2.397 & 9.588 \\ \hline
8 & 0.603 & 2.413 & 9.654 & 38.618 \\ \hline
16 & 2.017 & 8.067 & 32.270 & 129.081 \\ \hline
32 & 6.058 & 24.234 & 96.944 & 387.776 \\ \hline
64 & 16.976 & 67.909 & 271.656 & 1086.625 \\ \hline
128 & 45.290 & 181.179 & 724.772 & 2899.087 \\ \hline
256 & 116.499 & 466.042 & 1864.307 & 7457.229\\ \hline
\end{tabular}
\end{table}
}

\section{Proposed In-Memory Unary Sorting}
\label{sec_prop_unary}

In this section, we propose a novel method for sorting unary data in memory to avoid the overheads of off-memory processing in the unary systems. We first discuss the basic operation of sorting two unary bit-streams in memory and then elaborate on the design of a complete unary sorting network.

\ignore{
\begin{table}[]
\caption{Latency and Area of In-Memory Sorting}
\begin{tabular}{|c|c|c|c|}
\hline
Proposed Method & Data-width & \begin{tabular}[c]{@{}c@{}}Latency\\ (Cycles)\end{tabular} & \begin{tabular}[c]{@{}c@{}}Area\\ (\# of memristors)\end{tabular} \\ \hline
Binary Sorting & $n$ & X1 & X2 \\ \hline
Unary Sorting & $n$ & 3 & $6 \times 2^n$ \\ \hline
\end{tabular}
\end{table}
}

\subsection{Basic Unary Sorting Unit}
\label{Sec:Basic_Sorting_unary}

\begin{figure}[t]
\centering
\includegraphics[trim={0cm 0cm 0cm 0cm},width=3.1in]{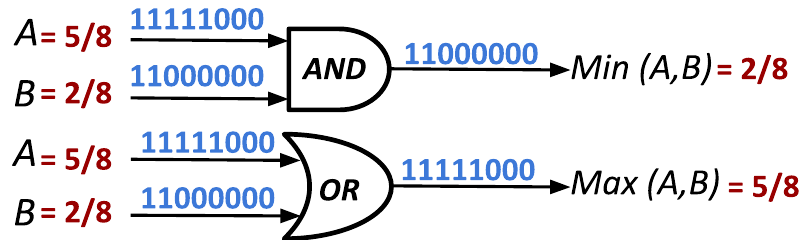}
\caption{Example of performing maximum and minimum operations on unary bit-streams.}
\label{fig:ExampleMaxMinUnary}
\end{figure}

The maximum and minimum functions are the essential operations in a basic sorting unit. Performing bit-wise logical \texttt{AND} on two unary bit-streams with the same length  gives the minimum of the two bit-stream. Bit-wise logical \texttt{OR}, on the other hand, gives the maximum of the two unary bit-streams with the same length. Fig.~\ref{fig:ExampleMaxMinUnary} shows an example of the maximum and the minimum operation on two unary bit-streams. The example presents these operations in a serial manner by processing one bit of the input bit-streams at each cycle. While the serial approach is extremely simple to implement with only one pair of \texttt{AND} and \texttt{OR} gates, it incurs a long latency proportional to the length of the bit-streams. In this work, we choose the precision of binary to unary conversion equal to the data-width, $n$. This means an $n$-bit data in the binary domain corresponds to a $2^n$-bit bit-stream in the unary domain. This implies a latency of $2^n$ cycles with a serial unit. Parallel sorting of two $n$-bit precision data represented using two $2^n$-bit bit-streams requires performing $2^n$ logical \texttt{AND} operations (to produce the minimum bit-stream), and $2^n$ logical \texttt{OR} operations (to produce the maximum bit-stream) in parallel as shown in Fig~\ref{Fig:CAS}(b). The suitability of the memristive crossbar for running parallel logical operations in memory makes it a perfect place for low-latency parallel sorting of unary bit-streams. 

\begin{figure}[t]
    \centering
\small (a)~\subfigure{
    \includegraphics[clip,trim={0.1cm 0cm 13cm 22cm}, width=2.5in]{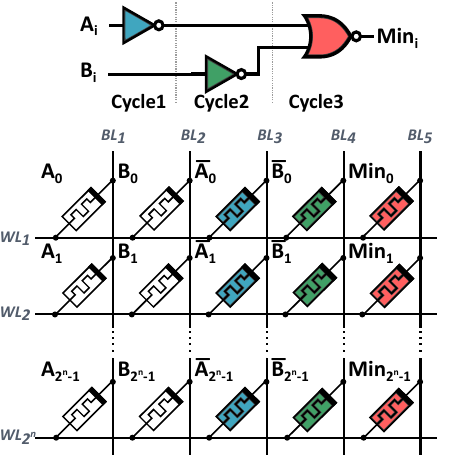}
\label{fig:Unary_MAGIC_Min}
}
\color{white}........\color{black}
\small ~~~(b)~\subfigure{
\includegraphics[clip,trim={0.1cm 0cm 15cm 22.5cm}, width=2in]{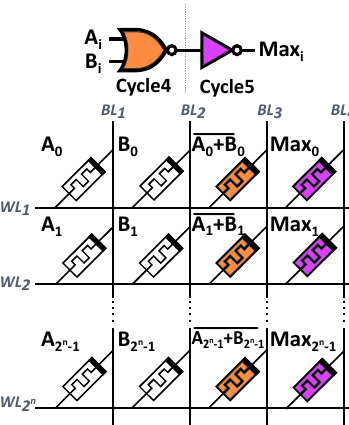}
\label{fig:Unary_MAGIC_Max}
}
\caption{Proposed in-memory unary sorting (a) in-memory minimum operation and (b) in-memory maximum operation on two unary bit-streams.(WL = Word Line, BL = Bit Line)}
\label{fig:UnaryMemory}
\end{figure}

\begin{table}[!t]
\centering
\caption{The Required Resources, Number of Processing Cycles, and Energy Consumption of the Proposed Basic Unary Sorting}
\label{Table:UnarySortingUnit}
\setlength{\tabcolsep}{3pt}
\vspace{-0.5em}
\renewcommand{\arraystretch}{1.1}
\resizebox{13.5cm}{!}
{
\begin{tabular}{|c|c|c|c|c|c|c|c|c|}
\hline
\begin{tabular}[c]{@{}c@{}}BitStream\\ Length\end{tabular} & \begin{tabular}[c]{@{}c@{}}Required\\ Dim.\end{tabular} & \begin{tabular}[c]{@{}c@{}}\# of Initialized\\ Memristors\end{tabular}  & \begin{tabular}[c]{@{}c@{}}\# of Reused\\ Memristors\end{tabular} & \begin{tabular}[c]{@{}c@{}}\# of \texttt{NOR}\\ Operations\end{tabular} & \begin{tabular}[c]{@{}c@{}}\# of \texttt{NOT}\\ Operations\end{tabular} & \begin{tabular}[c]{@{}c@{}}Initial\\ Cycles\end{tabular} & \begin{tabular}[c]{@{}c@{}}Operation \\ Cycles\end{tabular} & \begin{tabular}[c]{@{}c@{}}Energy \\ ($pJ$)\end{tabular}\\ \hline
$2^4$ & $16\times5$ & 64 & 32 & 32 & 48 & 1 & 5 & 227\\ \hline
$2^6$ & $64\times5$ & 256 & 128 & 128 & 192 & 1 & 5 & 910 \\ \hline
$2^8$ & $256\times5$ & 1024 & 512 & 512 & 768 & 1 & 5 & 3640 \\ \hline
$2^{10}$ & $1024\times5$ & 4096 & 2048 & 2048 & 3072 & 1 & 5 & 14558\\ \hline
\end{tabular}
}
\end{table}

Fig.~\ref{fig:UnaryMemory} shows our proposed design for MAGIC-based in-memory execution of  minimum and maximum operations on two unary bit-streams. As shown in Fig.~\ref{fig:UnaryMemory}, implementing this sorting unit using MAGIC \texttt{NOR} and \texttt{NOT} operations requires a memristor crossbar proportional to the  length  of  the  bit-streams. The unsorted unary data (i.e., A and B bit-streams) are stored in two different columns (BL$_1$ and BL$_2$). Both inputs  have the same length of $2^n$.  As shown in Fig.~\ref{fig:Unary_MAGIC_Min}, the \texttt{AND} operation (minimum function) is realized by first inverting the bit-streams through MAGIC \texttt{NOT} and then performing bit-wise MAGIC \texttt{NOR} on the inverted bit-streams. This effectively implements the \texttt{AND} operation as $A \land B = \overline{\overline{A}~{\lor}~ \overline{B}}$. The first and the second bit-stream are inverted on BL$_3$ and BL$_4$ in the first and the second cycle, respectively. The \texttt{NOR} operation is executed in the third cycle on BL$_5$. As shown in Fig.~\ref{fig:Unary_MAGIC_Max}, the \texttt{OR} operation (maximum function) is achieved by first performing MAGIC \texttt{NOR} on the input bit-streams and then MAGIC \texttt{NOT} on the outputs of the \texttt{NOR} operations. Hence, the execution of the \texttt{OR} operation takes two cycles. 

The columns that we use during the execution of the \texttt{AND} operation to store the inverted version of the bit-streams (e.g., the third and fourth columns in Fig.~\ref{fig:Unary_MAGIC_Min}) are re-used in the execution of the \texttt{OR} operation to avoid using additional memristors. In contrast to the proposed in-memory binary sorting of Section~\ref{Sec:Basic_Sorting_binary}, which has a variable latency dependent on the width of the input data, \textit{the processing latency of the proposed unary sorting is fixed} at five cycles and does not change with the data-width.
Table~\ref{Table:UnarySortingUnit} shows the required resources, number of cycles, and energy consumption of the proposed basic sorting unit for different bit-stream lengths. 

The number of memristors is directly proportional to the length of the bit-streams. In a fully parallel design approach, the size of the memory, particularly the number of rows, defines an upper-limit on the maximum data-width for the to-be-sorted unary data. In such a system, bit-streams with a length longer than the number of rows can be supported by splitting each bit-stream into multiple shorter sub-bit-streams, storing each sub-bit-stream in a different column, 
and executing the \gls{cas} operations in parallel. The  sub-results will be finally merged to produce the complete minimum and maximum bit-streams. This design approach sorts the data with reduced latency as the primary objective. A different approach for sorting long bit-streams is to perform \gls{cas} operations on the sub-bit-streams in a serial manner by re-using the CAS unit(s). 
The above approach reduces the area (number of used memristors) at the cost of additional latency. In this case, after sorting each pair of  sub-bit-streams, the result is saved, and a new pair of sub-bit-stream is loaded for sorting. Assuming that each input bit-stream is split into $N$ sub-bit-streams, the number of processing cycles to sort each pair of input data increases by a factor of $N$. Some additional processing cycles are also needed for saving each sub-output and copying each pair of sub-input. 
Combining the parallel and the serial approach is also possible for further trade-offs between area and delay. These approaches increase the range of supported data-widths but incur a more complicated implementation and partition management.

\subsection{Complete Unary Sort System}

Implementing a  bitonic sorting network in the unary domain follows the same approach as presented in Section~\ref{Sec:Binary-Complete-sort} for binary implementation of sorting networks. The number of sorting steps and the required number of basic sorting operations are exactly the same as those of the binary sorting network design. The essential difference, however,  is that in the unary sorting system, the data is in the unary format. Therefore, the basic 2-input sorting operation should be implemented based on the unary sorting unit proposed in Section~\ref{Sec:Basic_Sorting_unary}. Table~\ref{Table:SortingNetwork-Lateny-Dimension} shows the number of processing cycles and the required size of memory for implementing unary bitonic networks of different sizes. We report the latency, area, and energy of these networks as well.

\section{Comparison and Application}
\label{sec_comp}

\subsection{Circuit-Level Simulations}
\label{Sec:CirSimulation}
We implemented a 16$\times$16 crossbar and necessary control~signals~in Cadence Virtuoso for circuit-level evaluation of the proposed designs. For memristor simulations, we used the Voltage Controlled  ThrEshold Adaptive Memristor (VTEAM) model~{\cite{VTEAM}}. The  Parameters  used  for  the VTEAM model can be seen in Table~\ref{tab:vteam_values}. We evaluated the designs in an analog mixed-signal environment by using the Spectre simulation platform with 0.1$ns$ transient step. For MAGIC operations, we applied $V_{SET}$=$2.08V$ with 1$ns$ pulse-width to initialize the output memristors to \gls{lrs}. For the simplicity of controller design, we consider the clock cycle period of 1.25$ns$ and $V_0$ pulse-width of 1$ns$ for all operations. $V_0$ voltage for \texttt{NOT}, 2-input \texttt{NOR}, 3-input \texttt{NOR}, and 4-input \texttt{NOR} is $ 1.1V$, $950mV$, $1.05V$, and $1.15V$, respectively. We perform the copy operations by using two consecutive \texttt{NOT} operations.

To estimate the total energy of in-memory computations, we first find the energy consumption of each operation. The energy number measured for each operation depends on the states of input memristors (i.e., LRS, HRS). We consider all possible cases when measuring the energy of each operation. For example, the 3-input \texttt{NOR} has eight possible combinations of input states. We consider the average energy of these eight cases as the energy of 3-input \texttt{NOR}. The average measured energy of different operations is reported in Table~\ref{tab:energy_values}. Note that higher energy consumption for \texttt{NOT} operation compared to 2-input \texttt{NOR} is due to using a higher $V_0$ voltage for \texttt{NOT}. The reported energy for the proposed in-memory sorting designs is the sum of the energy consumed by all operations.

\begin{table}[]
\centering
\caption{\label{tab:vteam_values}
Memristor Parameter Values from~\cite{MAGIC_2014} for the VTEAM Model~\cite{VTEAM}.
}

\resizebox{3in}{!}
{
\begin{tabular}{|cccc|}
\hline
Parameter & \multicolumn{1}{c|}{{Value}} & {Parameter} & {Value} \\ \hline
{$R_{on}$} & \multicolumn{1}{c|}{{1 k$\Omega$}} & {$x_{off}$} & {3 $nm$} \\
$R_{off}$ & \multicolumn{1}{c|}{300 k$\Omega$} & $k_{on}$ & -216.2 m/sec \\
$VT_{on}$ & \multicolumn{1}{c|}{-1.5 V} & $k_{off}$ & 0.091 m/sec \\
$VT_{off}$ & \multicolumn{1}{c|}{300 mV} & $\alpha_{on}$ & 4 \\
$x_{on}$ & \multicolumn{1}{c|}{0 nm} & $\alpha_{off}$ & 4 \\ \hline
\end{tabular}
}
\end{table}

\begin{table}[]

\centering
\caption{\label{tab:energy_values}{The Average Measured Energy Consumption of Each Operation Based on VTEAM Model.}}

{
\resizebox{2.5in}{!}{
\begin{tabular}{|c|c|}
\hline
Operation      & Average Energy \\ \hline
memristor initialization & 2350 $fJ$         \\ \hline
memristor copy           & 40.08 $fJ$        \\ \hline
\texttt{NOT}            & 20.04 $fJ$        \\ \hline
2-input \texttt{NOR}    & 9.01 $fJ$         \\ \hline
3-input \texttt{NOR}    & 37.24 $fJ$        \\ \hline
4-input \texttt{NOR}    & 54.51 $fJ$        \\ \hline
\end{tabular}
}}
\vspace{-1em}
\end{table}

\subsection{Comparison of In- and Off-Memory}

We compare the latency and energy consumption of the proposed in-memory binary and unary sorting designs with the conventional off-memory CMOS-based designs for the case of implementing bitonic networks with a data-width of eight. For a fair comparison, we assume that the to-be-sorted data are already stored in memristive memory when the sorting process  begins and hence do  not consider the delay for initial storage. We do  not consider this latency  because it is the same  for both cases of the  proposed in-memory and  the off-memory counterparts. For the case of off-memory binary designs, we assume 8-bit precision data are read from and written to a memristive memory. For the case of off-memory unary design, we evaluate two approaches: 1)~unary data (i.e., 256-bit bit-streams) are read from and written to memory, and 2) 8-bit binary data are read from and written to memory. For the second approach, the conversion overhead (i.e., binary to/from unary bit-stream) is also considered. This conversion is performed off-memory using combinational CMOS logic~\cite{Sorting-TVLSI-2018}. The  conventional CMOS-based  off-memory sorting systems read the raw data from memory, sort the data with CMOS logic, and write the sorted data into  memory. These  read and  write  operations take the largest portion of the latency and energy consumption. 
We use the per-bit read and write latency and per-bit energy consumption reported in~\cite{DATE_2011_CongXu} to calculate the total latency and energy of reading from and writing into the memristive memory.
For the proposed in-memory designs, the  entire  processing step is  performed in memory,  and  so there is no read  and  write operations from and to the memory. 
For the off-memory cases, we do not incorporate the transferring overhead between the memory and the processing unit as it depends on the interconnects used.
We implemented the off-memory processing units using Verilog HDL and synthesized them using the Synopsys Design Compiler v2018.06-SP2 with the 45nm NCSU-FreePDK gate library.

Table~\ref{Table:EnergyLatencyComp} shows the summary of performance results. As reported, the proposed in-memory designs provide a significant latency and energy reduction, compared to the conventional off-memory designs. That is, on average 14$\times$ and 37$\times$, respectively, for the binary sorting. For the unary design, the average latency and energy reductions are 1200$\times$ and 138$\times$, respectively. For the unary systems with the data stored in memory in a binary
format, the proposed in-memory design can reduce the latency and energy by a factor of up to 65$\times$ and 9.7$\times$, respectively. For a realistic and more accurate energy consumption comparison, however, the overhead of transferring data on the interconnect between the memory and the processing unit must be added for the off-memory cases. We note that these numbers are highly dependent on the architecture of the overall system and the interconnects used. Therefore, different system architectures may substantially change these numbers; however, they do not change the fact that our proposed method is more advantageous. In fact, they only change the extent of this improvement (and further increase it) since no data transfer happens in the in-memory sorting solution. Hence, by eliminating them, we present the minimum improvement obtained by our method and leave the further improvement to the final implementation details of designers.

\begin{table}[!t]

\caption{{Energy Consumption ($nJ$) and Latency ($\mu s$) of the Implemented In-Memory and Off-Memory Bitonic Sorting Designs with Data-Width=8} (E: Energy, L: Latency)}
\centering
\label{Table:EnergyLatencyComp}
\setlength{\tabcolsep}{3pt}
\vspace{-0.75em}
\renewcommand{\arraystretch}{1.2}
\resizebox{16cm}{!}
{
\begin{tabular}{|c|c|c|c|c|c|c|c|c|c|c|c|c|}
\hline
Network Size & \multicolumn{2}{c|}{8} & \multicolumn{2}{c|}{16} & \multicolumn{2}{c|}{32} & \multicolumn{2}{c|}{64} & \multicolumn{2}{c|}{128} & \multicolumn{2}{c|}{256} \\ \hline
Design Method & E & L & E & L & E & L & E & L & E & L & E & L \\ \hline
Off-Memory Binary Sorting (+ Binary R/W) & 850 & 6.5 & 1701 & 13 & 3403 & 26 & 6806 & 52 & 13613 & 104 & 27227 & 209 \\ \hline
 Proposed In-Memory Binary Sorting & 10 & 0.55 & 33 & 1.02 & 100 & 1.8 & 281 & 3.4 & 794 & 6.8 & 1927 & 14  \\ \hline \hline
Off-Memory Unary Sorting (+ Binary R/W) & 851 & 6.5 & 1703 & 13 & 3406 & 26 & 6811 & 52 & 13622 & 104 & 27244 & 209 \\ \hline
Off-Memory Unary Sorting (+ Unary R/W) & 27226 & 210 & 54452 & 419 & 108904 & 839 & 217809 & 1679 & 435618 & 3358 & 871236 & 6717 \\ \hline
 Proposed In-Memory Unary Sorting & 87 & 0.10 & 291 & 0.25& 875 & 0.7& 2452 &1.8& 6,540 &4.7& 16,817 &12 \\ \hline
\end{tabular}
}

\end{table}

\ignore{
\begin{table}[h]
\centering
\caption{\blue{Hardware Area ($\mu m^2$) Cost of the Implemented Off-Memory  Sorting Designs and the Area Overhead of the Peripheral Circuits for the In-Memory Sorting Architectures. The Data-Width is 8.}}
\vspace{-1.5em}
\label{Table:AreaComp}
\blue{
\begin{tabular}{ccccccc}

& \multicolumn{6}{c}{} \\ \hline
\multicolumn{1}{|c|}{Network size} & \multicolumn{1}{c|}{8} & \multicolumn{1}{c|}{16} & \multicolumn{1}{c|}{32} & \multicolumn{1}{c|}{64} & \multicolumn{1}{c|}{128} & \multicolumn{1}{c|}{256} \\ \hline
\multicolumn{1}{|c|}{In-memory Binary Sorting} & \multicolumn{1}{c|}{11301} & \multicolumn{1}{c|}{23471} & \multicolumn{1}{c|}{48682} & \multicolumn{1}{c|}{100842} & \multicolumn{1}{c|}{208638} & \multicolumn{1}{c|}{403368} \\ \hline
\multicolumn{1}{|l|}{Off-Memory Binary Sorting (+ Binary R/W)} & \multicolumn{1}{c|}{6926} & \multicolumn{1}{c|}{19338} & \multicolumn{1}{c|}{57900} & \multicolumn{1}{c|}{161934} & \multicolumn{1}{c|}{431062} & \multicolumn{1}{c|}{1107998} \\ \hline
 \hline
\multicolumn{1}{|c|}{In-memory Unary Sorting} & \multicolumn{1}{c|}{642} & \multicolumn{1}{c|}{1326} & \multicolumn{1}{c|}{2739} & \multicolumn{1}{c|}{5649} & \multicolumn{1}{c|}{11641} & \multicolumn{1}{c|}{23968} \\ \hline
\multicolumn{1}{|c|}{Off-Memory Unary Sorting (+ Binary R/W)} & \multicolumn{1}{c|}{2659} & \multicolumn{1}{c|}{5834} & \multicolumn{1}{c|}{13095} & \multicolumn{1}{c|}{25248} & \multicolumn{1}{c|}{59579} & \multicolumn{1}{c|}{140006} \\ \hline
\end{tabular}
}
\end{table}
}

\subsection{Application to Median Filtering}

Median filtering has been widely used in different applications, from image and video to speech and signal processing. In these applications, digital data is often affected by noise. A median filter ---which replaces each input data with the median of all the data in a local neighborhood (e.g., a 3$\times$3 local window)--- is used to filter out impulse noises and smoothen the data~\cite{MedianPaper3}. A variety of methods for the implementation of Median filters have been proposed. Sorting network-based architectures made of \gls{cas} blocks are one of the most common approaches~\cite{Sorting-TVLSI-2018}. The incoming data is sorted as it passes the network. The middle element of the sorted data is the median. We developed in-memory architectures for a $3\times3$ and a $5\times5$ median filtering based on our proposed in-memory binary and unary sorting designs. 

\begin{figure}[!t]
\centering
\includegraphics[clip,trim={0.0cm 0cm 5.3cm 18.9cm},width=3.0in]{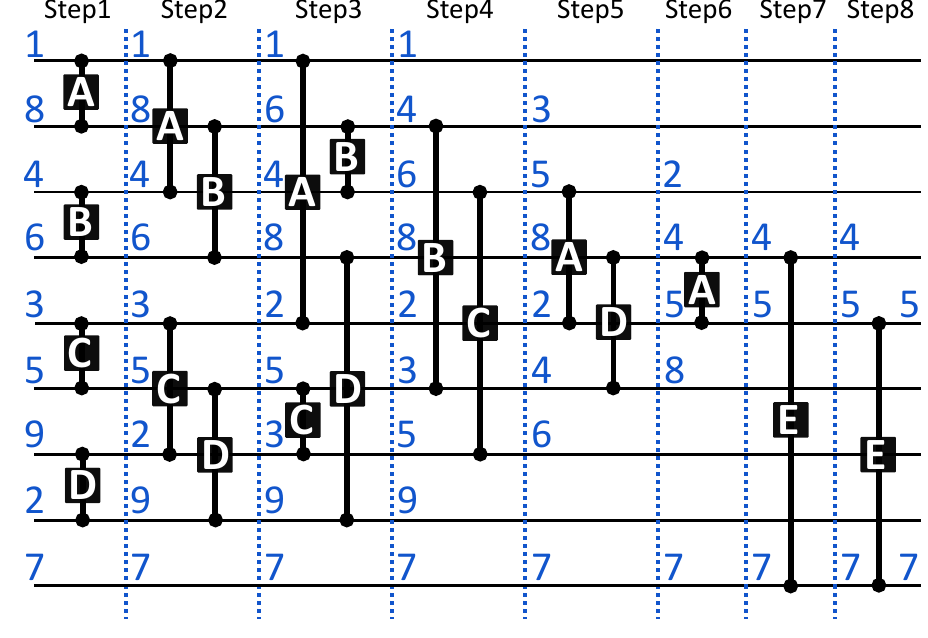}
\caption{Processing Steps and Memory Partitioning of the 3$\times$3 Median Filter Design.
}
\label{fig:Medianfilter3x3}
\end{figure}

\begin{figure}[!t]
\centering
\includegraphics[clip,trim={0.0cm 10.5cm 0cm 0.0cm},width=7.in]{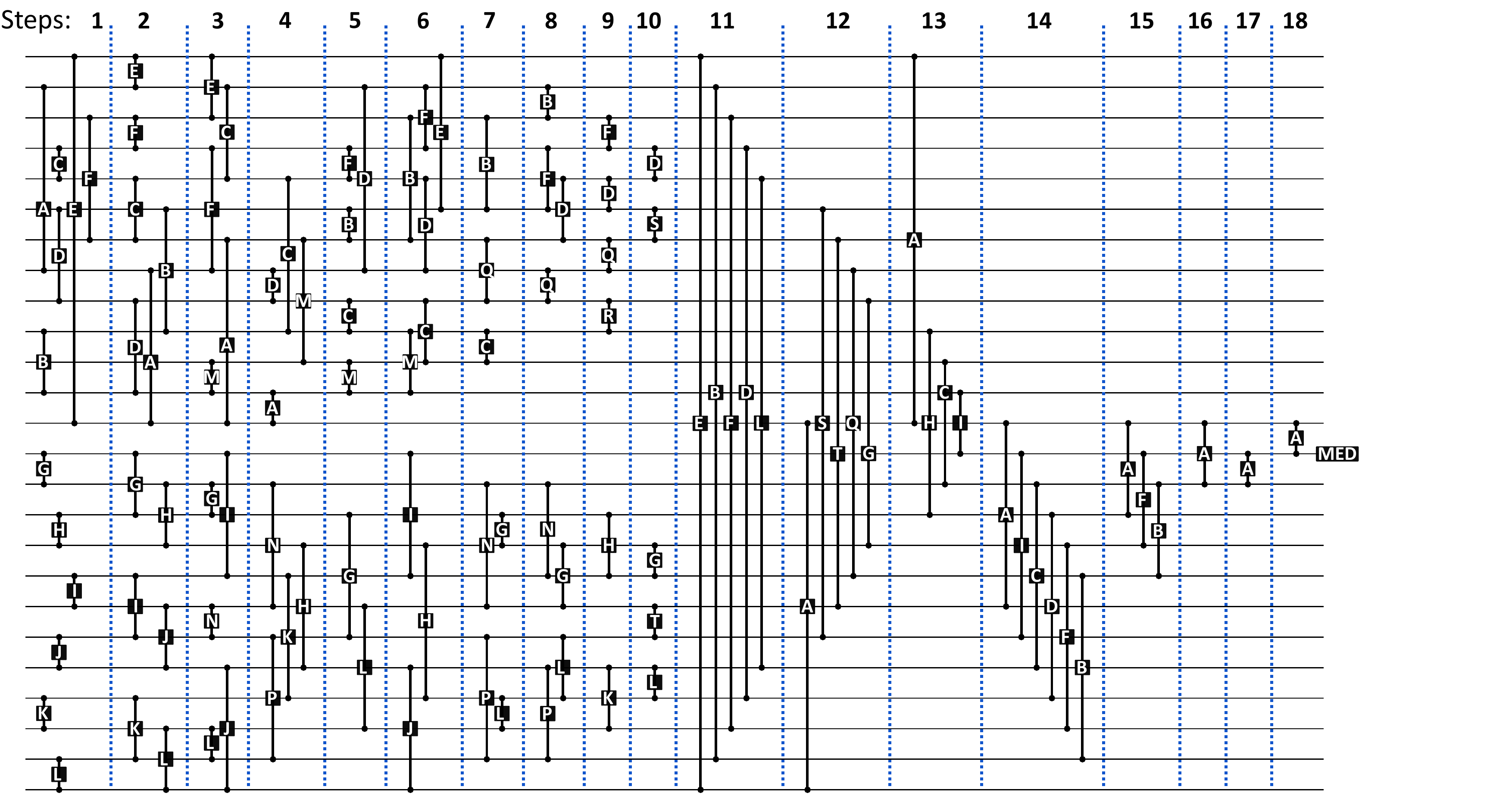}
\vspace{-1.25em}
\caption{Processing Steps and Memory Partitioning of the 5$\times$5 Median Filter Design.
}
\vspace{-0.5em}
\label{fig:Medianfilter5x5}
\end{figure}

Fig.~\ref{fig:Medianfilter3x3} and Fig.~\ref{fig:Medianfilter5x5} depict a high-level flow of memory partitioning for our in-memory $3\times3$ and $5\times5$ Median filter design. Similar to our approach in implementing the complete sort system, the memory is split into multiple partitions. For the $3\times3$ design, partitions are A, B, C, D, and E (five partitions), and for the $5\times5$ design, they are A to T (20 partitions). Each sorting unit sorts the data in one particular partition.
Some partitions are initialized with the input data in the first step and the others are initialized and used in the following steps. The process is split into eight steps for the $3\times3$, and to 18 steps for the $5\times5$ design, each step executing some basic sorting operations in parallel. After each step, to prepare data for the next step, some data must be transferred between partitions similar to what we did in sorting. These data transfers are done by using copy operations. Compared to a complete sorting network, fewer sorting units are required as only the median value is targeted.

We evaluated the implementation of these basic sorting operations using both the proposed binary and unary bit-stream-based in-memory architectures.
Table~\ref{Table:MedianFilter} reports the latency, the number of required memristors, and the energy consumption of the developed designs for (i)~a single $3\times3$ Median filter and a single $5\times5$ Median filter and (ii) a $3\times3$ and a $5\times5$ Median filter image processing system that process images of $64\times64$ size. The corresponding latency and energy consumption of the off-memory CMOS-based binary and unary designs are also reported in Table~\ref{Table:MedianFilter}. As it can be seen, the proposed in-memory binary and unary designs reduce the energy by a factor of 14$\times$ and 5.6$\times$, respectively, for the $3\times3$-based image processing system, and 3.1$\times$ and 12$\times$ for the $5\times5$-based
image processing system, compared to their corresponding off-memory designs. The latency of the binary and unary design is also reduced by a factor of  634$\times$ and 152$\times10^3$ $\times$ with the $3\times3$ window, and by a factor of 110$\times$ and 19.2$\times10^3$ $\times$ with the $5\times5$ window, for the $64\times64$ image processing system.

Note that we did not incorporate the overhead latency and the energy of transferring data on the bus or other interconnects for the off-memory cases, which is a large portion of energy of consumption in transferring data between memory and processing unit~\cite{memory_energy_wire}. By considering this overhead, our approach would have a significantly larger advantage over others in a complete system.

\begin{table}[]
\centering
\caption{The Required Resources {($M_t$)}, Latency (L), and Energy Consumption (E) of the Implemented Median Filter Designs}
\vspace{-0.5em}
\label{Table:MedianFilter}
\setlength{\tabcolsep}{2pt}
\renewcommand{\arraystretch}{1.2}
\resizebox{4.8in}{!}{
\begin{tabular}{|c|c|c|c|c||c|c|c|}
\hline
 & \multicolumn{7}{c|}{Median Filter 3 $\times$ 3} \\ \hline
Design & 
{$PC_t$}
& 
{$M_t$}
& E ($\mu$ J) & L ($\mu$ s) & Design & E ($\mu$ J) & L ($\mu$ s) \\ \hline
Proposed Binary & 544 & 8 $\times$ 110 & 0.0085 & 0.68 & Off-Memory Binary & 0.121 & 0.94 \\ \hline
Proposed Unary & 72 & 256 $\times$ 25 & 0.069 & 0.09 & Off-Memory Unary & 3.882 & 30.17 \\ \hline\hline
 & \multicolumn{7}{c|}{64 $\times$ 64 Image Processor} \\ \hline
Proposed Binary & 4896 & 208 $\times$ 1980 & 35 & 6.1 & Off-Memory Binary & 490 & 3870 \\ \hline
Proposed Unary & 684 & 2048 $\times$ 1425 & 283 & 0.81 & Off-Memory Unary & 1590 & 123578 \\ \hline
\end{tabular}
}

\resizebox{4.8in}{!}{
\begin{tabular}{|c|c|c|c|c||c|c|c|}
\hline
 & \multicolumn{7}{c|}{Median Filter 5 $\times$ 5} \\ \hline
Design & 
{$PC_t$}
& 
{$M_t$}
& E ($\mu$ J) & L ($\mu$ s) & Design & E ($\mu$ J) & L ($\mu$ s) \\ \hline
Proposed Binary & 1416 & 8 $\times$ 440 & 0.049 & 1.77 & Off-Memory Binary & 0.151 & 1.19 \\ \hline
Proposed Unary & 259 & 256 $\times$ 100 & 0.401 & 0.324 & Off-Memory Unary & 4841 & 38.02 \\ \hline\hline
 & \multicolumn{7}{c|}{64 $\times$ 64 Image Processor} \\ \hline
Proposed Binary & 35400 & 328 $\times$ 1760 & 200 & 44.25 & Off-Memory Binary & 620 & 4875 \\ \hline
Proposed Unary & 6475 & 2048 $\times$ 2000 & 1643 & 8.09 & Off-Memory Unary & 19829 & 155739 \\ \hline
\end{tabular}
}
\vspace{-2em}

\end{table}

\ignore{
\begin{tabular}{|c|c|c|c|c||c|c|c|c|}
\hline
\multirow{2}{*}{Design} & \multicolumn{4}{c||}{Median Filter $3\times3$} & \multicolumn{4}{c|}{$64\times64$ Image Processor} \\ \cline{2-9} 
 & Cycle & Area & E ($\mu J$) & L ($\mu s$) & Cycle & Area & E ($\mu J$) & L ($\mu s$) \\ \hline
\begin{tabular}[c]{@{}c@{}}Proposed\\ Binary\end{tabular} & 544 & $8\times110$ & 0.0085 & 0.68 & 4896 & $208\times1980$ & 35 & 6.1 \\ \hline
\begin{tabular}[c]{@{}c@{}}Proposed\\ Unary\end{tabular} & 72 & $256\times25$ & 0.069 & 0.09 & 684 & $2048\times1425$ & 283 & 0.81 \\ \hline
\end{tabular}
\begin{tabular}{|c|c|c||c|c|}
\hline
\multirow{2}{*}{Design} & \multicolumn{2}{c||}{Median Filter $3\times3$} & \multicolumn{2}{c|}{ $64\times64$ Image Processor} \\ \cline{2-5} 
 & E ($\mu J$) & L ($\mu s$) & E ($\mu J$) & L ($\mu s$) \\ \hline
\begin{tabular}[c]{@{}c@{}}Off-Memory\\ Binary\end{tabular}& 0.121 & 0.94 & 490 & 3200 \\ \hline
\begin{tabular}[c]{@{}c@{}}Off-Memory\\ Unary\end{tabular} & 3.882 & 30 & 1590 & 104000 \\ \hline
\end{tabular}
}

\section{Discussion}
\label{sec_discussion}
The high latency and energy overhead of reading from and writing to memory, and transferring data between  the processing unit and memory, take up a significant amount of resources in sorting data in the conventional systems. \gls{imc} is a promising solution to mitigate these overheads. \gls{imc} is particularly beneficial for 1) applications with large data or a large number of memory accesses and 2) applications with extensive parallelism that can independently run a large number of operations in parallel. Sorting is one of the applications that has both properties. As we showed, implementing sorting in memory can save significant time and energy by avoiding the overheads of memory access and off-chip data transfer. This is particularly important for the unary systems, where data are stored in memory in the form of long bit-streams. Reading and writing long bit-streams from and to memory make off-memory unary sorting highly inefficient.
However, one should note that the size of the memory array puts an upper limit on the size of the sorting network and the data-width. For example, given a (memristive) memory array of $1024\times1024$, the proposed binary sorting approach supports the complete sorting of 64 8-bit and 128 4-bit input data. For the proposed unary approach, an array of that size (1Mb or 128kB) supports the complete sorting of 256 1024-bit unary bit-streams. For the larger bit-streams or data-width, or larger number of data to be sorted, we would need to partition the data into different arrays. That means a more complex control and partition management mechanism, which reduces the benefits of fully in-array sorting.

Memristive technology is an emerging technology still in evolution, with many competing implementation methods in the process of maturation~\cite{Taherinejad2019cas}. Properties such as the delay, power, and energy consumption are heavily dependent on the used technology and change considerably from one to another. In this article, the experimental results are provided by simulation tools using the VTEAM model. Using different memristive technologies and models, LRS and HRS values, as well as programming or reading pulses with different amplitude or width, affect the actual delay and energy numbers reported here. The actual memristive implementation is a showcase of the feasibility and meaningfulness of such an in-memory sorting design. That is, there exists an in-memory implementation (namely using the memristive technology we have used here) to be significantly beneficial. Therefore, we consider the properties and comparison of other ways of implementing our proposed architecture (using CMOS \gls{imc}, other memristor technologies, or other emerging memory technologies) as an exciting future work but outside the current article's scope. Nonetheless, we would like to point out that we have provided the number of memristors (memory cells) and the number of operation cycles that are technology-independent. Therefore, others can independently evaluate and compare their own implementations with ours in a technology-agnostic fashion and using the number of memory cells and the number of cycles their implementation needs (regardless of the LRS and HRS values or pulse amplitude and width).

\section{Conclusion}
\label{sec_conclusions}

Thus far, sorting solutions were based on the conventional approach of processing off-memory, incurring a high overhead of reading/writing from/to memory and transferring between the memory and the processing unit. 
In this paper, for the first time -to the best of our knowledge- we developed two methods for in-array sorting of data: a binary and a unary sorting design. We compared the area, latency, and energy consumption of the basic and the complete sorting systems for different data-widths and network sizes. The latency and energy are significantly reduced compared to prior off-memory CMOS-based sorting designs. Further, we developed in-memory binary and unary designs for an important sorting application, median filtering. In future works, we plan to extend the proposed architectures to other applications of sorting, for instance, efficient in-memory implementation of weighted and adaptive median filters. We will also explore applications of in-memory sorting in communications and coding domains.

\bibliographystyle{ACM-Reference-Format}
\bibliography{references, Hassan, nima}


\begin{thebibliography}{69}


\ifx \showCODEN    \undefined \def \showCODEN     #1{\unskip}     \fi
\ifx \showDOI      \undefined \def \showDOI       #1{#1}\fi
\ifx \showISBNx    \undefined \def \showISBNx     #1{\unskip}     \fi
\ifx \showISBNxiii \undefined \def \showISBNxiii  #1{\unskip}     \fi
\ifx \showISSN     \undefined \def \showISSN      #1{\unskip}     \fi
\ifx \showLCCN     \undefined \def \showLCCN      #1{\unskip}     \fi
\ifx \shownote     \undefined \def \shownote      #1{#1}          \fi
\ifx \showarticletitle \undefined \def \showarticletitle #1{#1}   \fi
\ifx \showURL      \undefined \def \showURL       {\relax}        \fi
\providecommand\bibfield[2]{#2}
\providecommand\bibinfo[2]{#2}
\providecommand\natexlab[1]{#1}
\providecommand\showeprint[2][]{arXiv:#2}

\bibitem[\protect\citeauthoryear{Aga, Jeloka, Subramaniyan, Narayanasamy,
  Blaauw, and Das}{Aga et~al\mbox{.}}{2017}]%
        {EqualityComparator1}
\bibfield{author}{\bibinfo{person}{Shaizeen Aga}, \bibinfo{person}{Supreet
  Jeloka}, \bibinfo{person}{Arun Subramaniyan}, \bibinfo{person}{Satish
  Narayanasamy}, \bibinfo{person}{David Blaauw}, {and}
  \bibinfo{person}{Reetuparna Das}.} \bibinfo{year}{2017}\natexlab{}.
\newblock \showarticletitle{Compute Caches}. In \bibinfo{booktitle}{\emph{2017
  IEEE International Symposium on High Performance Computer Architecture
  (HPCA)}}. \bibinfo{pages}{481--492}.
\newblock
\urldef\tempurl%
\url{https://doi.org/10.1109/HPCA.2017.21}
\showDOI{\tempurl}


\bibitem[\protect\citeauthoryear{Al-Haj~Baddar and Mahafzah}{Al-Haj~Baddar and
  Mahafzah}{2014}]%
        {Review2Ref1}
\bibfield{author}{\bibinfo{person}{Sherenaz~W. Al-Haj~Baddar} {and}
  \bibinfo{person}{Basel~A. Mahafzah}.} \bibinfo{year}{2014}\natexlab{}.
\newblock \showarticletitle{Bitonic Sort on a Chained-cubic Tree
  Interconnection Network}.
\newblock \bibinfo{journal}{\emph{JPDC}} \bibinfo{volume}{74},
  \bibinfo{number}{1} (\bibinfo{year}{2014}), \bibinfo{pages}{1744--1761}.
\newblock
\showISSN{0743-7315}
\urldef\tempurl%
\url{https://doi.org/10.1016/j.jpdc.2013.09.008}
\showDOI{\tempurl}


\bibitem[\protect\citeauthoryear{Alaghi and Hayes}{Alaghi and Hayes}{2013}]%
        {alaghi_Correlation}
\bibfield{author}{\bibinfo{person}{Armin Alaghi} {and} \bibinfo{person}{John~P.
  Hayes}.} \bibinfo{year}{2013}\natexlab{}.
\newblock \showarticletitle{Exploiting correlation in stochastic circuit
  design}. In \bibinfo{booktitle}{\emph{2013 IEEE 31st International Conference
  on Computer Design (ICCD)}}. \bibinfo{pages}{39--46}.
\newblock
\urldef\tempurl%
\url{https://doi.org/10.1109/ICCD.2013.6657023}
\showDOI{\tempurl}


\bibitem[\protect\citeauthoryear{Alaghi, Qian, and Hayes}{Alaghi
  et~al\mbox{.}}{2018}]%
        {Armin_Survey_2018}
\bibfield{author}{\bibinfo{person}{Armin Alaghi}, \bibinfo{person}{Weikang
  Qian}, {and} \bibinfo{person}{John~P. Hayes}.}
  \bibinfo{year}{2018}\natexlab{}.
\newblock \showarticletitle{The Promise and Challenge of Stochastic Computing}.
\newblock \bibinfo{journal}{\emph{IEEE Transactions on Computer-Aided Design of
  Integrated Circuits and Systems}} \bibinfo{volume}{37}, \bibinfo{number}{8}
  (\bibinfo{year}{2018}), \bibinfo{pages}{1515--1531}.
\newblock
\urldef\tempurl%
\url{https://doi.org/10.1109/TCAD.2017.2778107}
\showDOI{\tempurl}


\bibitem[\protect\citeauthoryear{Alam, Hassan~Najafi, and Nejad}{Alam
  et~al\mbox{.}}{2021}]%
        {Riahi_DesignTest21}
\bibfield{author}{\bibinfo{person}{Mohsen~Riahi Alam}, \bibinfo{person}{M.
  Hassan~Najafi}, {and} \bibinfo{person}{Nima~Taheri Nejad}.}
  \bibinfo{year}{2021}\natexlab{}.
\newblock \showarticletitle{Exact Stochastic Computing Multiplication in
  Memristive Memory}.
\newblock \bibinfo{journal}{\emph{IEEE Design Test}} (\bibinfo{year}{2021}),
  \bibinfo{pages}{1--1}.
\newblock
\urldef\tempurl%
\url{https://doi.org/10.1109/MDAT.2021.3051296}
\showDOI{\tempurl}


\bibitem[\protect\citeauthoryear{Angizi, He, Rakin, and Fan}{Angizi
  et~al\mbox{.}}{2018}]%
        {Angizi_DAC19}
\bibfield{author}{\bibinfo{person}{Shaahin Angizi}, \bibinfo{person}{Zhezhi
  He}, \bibinfo{person}{Adnan~Siraj Rakin}, {and} \bibinfo{person}{Deliang
  Fan}.} \bibinfo{year}{2018}\natexlab{}.
\newblock \showarticletitle{CMP-PIM: An Energy-Efficient Comparator-Based
  Processing-in-Memory Neural Network Accelerator}. In
  \bibinfo{booktitle}{\emph{Proceedings of the 55th Annual Design Automation
  Conference}} (San Francisco, California) \emph{(\bibinfo{series}{DAC '18})}.
  \bibinfo{publisher}{Association for Computing Machinery},
  \bibinfo{address}{New York, NY, USA}, Article \bibinfo{articleno}{105},
  \bibinfo{numpages}{6}~pages.
\newblock
\showISBNx{9781450357005}
\urldef\tempurl%
\url{https://doi.org/10.1145/3195970.3196009}
\showDOI{\tempurl}


\bibitem[\protect\citeauthoryear{Balasubramonian, Chang, Manning, Moreno,
  Murphy, Nair, and Swanson}{Balasubramonian et~al\mbox{.}}{2014}]%
        {NDPMicro14}
\bibfield{author}{\bibinfo{person}{Rajeev Balasubramonian},
  \bibinfo{person}{Jichuan Chang}, \bibinfo{person}{Troy Manning},
  \bibinfo{person}{Jaime~H. Moreno}, \bibinfo{person}{Richard Murphy},
  \bibinfo{person}{Ravi Nair}, {and} \bibinfo{person}{Steven Swanson}.}
  \bibinfo{year}{2014}\natexlab{}.
\newblock \showarticletitle{Near-Data Processing: Insights from a MICRO-46
  Workshop}.
\newblock \bibinfo{journal}{\emph{IEEE Micro}} \bibinfo{volume}{34},
  \bibinfo{number}{4} (\bibinfo{year}{2014}), \bibinfo{pages}{36--42}.
\newblock
\urldef\tempurl%
\url{https://doi.org/10.1109/MM.2014.55}
\showDOI{\tempurl}


\bibitem[\protect\citeauthoryear{Batcher}{Batcher}{1968}]%
        {SortingBatcher}
\bibfield{author}{\bibinfo{person}{Ken~E. Batcher}.}
  \bibinfo{year}{1968}\natexlab{}.
\newblock \showarticletitle{Sorting Networks and Their Applications}. In
  \bibinfo{booktitle}{\emph{Proceedings of the April 30--May 2, 1968, Spring
  Joint Computer Conference}} (Atlantic City, New Jersey)
  \emph{(\bibinfo{series}{AFIPS '68 (Spring)})}.
  \bibinfo{publisher}{Association for Computing Machinery},
  \bibinfo{address}{New York, NY, USA}, \bibinfo{pages}{307–314}.
\newblock
\showISBNx{9781450378970}
\urldef\tempurl%
\url{https://doi.org/10.1145/1468075.1468121}
\showDOI{\tempurl}


\bibitem[\protect\citeauthoryear{Bey Ahmed~Khernache, Laga, and Boukhobza}{Bey
  Ahmed~Khernache et~al\mbox{.}}{2018}]%
        {Khernache17}
\bibfield{author}{\bibinfo{person}{Mohammed Bey Ahmed~Khernache},
  \bibinfo{person}{Arezki Laga}, {and} \bibinfo{person}{Jalil Boukhobza}.}
  \bibinfo{year}{2018}\natexlab{}.
\newblock \showarticletitle{MONTRES-NVM: An External Sorting Algorithm for
  Hybrid Memory}. In \bibinfo{booktitle}{\emph{2018 IEEE 7th Non-Volatile
  Memory Systems and Applications Symposium (NVMSA)}}. \bibinfo{pages}{49--54}.
\newblock
\urldef\tempurl%
\url{https://doi.org/10.1109/NVMSA.2018.00013}
\showDOI{\tempurl}


\bibitem[\protect\citeauthoryear{Borghetti, Snider, Kuekes, Yang, Stewart, and
  Stanley~Williams}{Borghetti et~al\mbox{.}}{2010}]%
        {borghetti2010nature}
\bibfield{author}{\bibinfo{person}{Julien Borghetti},
  \bibinfo{person}{Gregory~S. Snider}, \bibinfo{person}{Philip~J. Kuekes},
  \bibinfo{person}{J.~Joshua Yang}, \bibinfo{person}{Duncan~R. Stewart}, {and}
  \bibinfo{person}{Richard Stanley~Williams}.} \bibinfo{year}{2010}\natexlab{}.
\newblock \showarticletitle{‘{Memristive}’ switches enable ‘stateful’
  logic operations via material implication}.
\newblock \bibinfo{journal}{\emph{Nature}}  \bibinfo{volume}{464}
  (\bibinfo{date}{April} \bibinfo{year}{2010}), \bibinfo{pages}{873--876}.
\newblock
\urldef\tempurl%
\url{https://doi.org/10.1038/nature08940}
\showDOI{\tempurl}


\bibitem[\protect\citeauthoryear{Brajovic and Kanade}{Brajovic and
  Kanade}{1999}]%
        {ApplicationRobotic}
\bibfield{author}{\bibinfo{person}{Vladimir Brajovic} {and}
  \bibinfo{person}{Takeo Kanade}.} \bibinfo{year}{1999}\natexlab{}.
\newblock \showarticletitle{A VLSI sorting image sensor: global massively
  parallel intensity-to-time processing for low-latency adaptive vision}.
\newblock \bibinfo{journal}{\emph{IEEE Transactions on Robotics and
  Automation}} \bibinfo{volume}{15}, \bibinfo{number}{1}
  (\bibinfo{year}{1999}), \bibinfo{pages}{67--75}.
\newblock
\urldef\tempurl%
\url{https://doi.org/10.1109/70.744603}
\showDOI{\tempurl}


\bibitem[\protect\citeauthoryear{Capannini, Silvestri, and Baraglia}{Capannini
  et~al\mbox{.}}{2012}]%
        {Sorting_GPU3}
\bibfield{author}{\bibinfo{person}{Gabriele Capannini},
  \bibinfo{person}{Fabrizio Silvestri}, {and} \bibinfo{person}{Ranieri
  Baraglia}.} \bibinfo{year}{2012}\natexlab{}.
\newblock \showarticletitle{{Sorting on GPUs for large scale datasets: A
  thorough comparison}}.
\newblock \bibinfo{journal}{\emph{Information Processing \& Management}}
  \bibinfo{volume}{48}, \bibinfo{number}{5} (\bibinfo{year}{2012}),
  \bibinfo{pages}{903 -- 917}.
\newblock
\showISSN{0306-4573}
\urldef\tempurl%
\url{https://doi.org/10.1016/j.ipm.2010.11.010}
\showDOI{\tempurl}
\newblock
\shownote{Large-Scale and Distributed Systems for Information Retrieval.}


\bibitem[\protect\citeauthoryear{Cederman and Tsigas}{Cederman and
  Tsigas}{2010}]%
        {Sorting_GPU1}
\bibfield{author}{\bibinfo{person}{Daniel Cederman} {and}
  \bibinfo{person}{Philippas Tsigas}.} \bibinfo{year}{2010}\natexlab{}.
\newblock \showarticletitle{{GPU-Quicksort: A Practical Quicksort Algorithm for
  Graphics Processors}}.
\newblock \bibinfo{journal}{\emph{ACM JEA}}  \bibinfo{volume}{14}, Article
  \bibinfo{articleno}{4} (\bibinfo{year}{2010}), \bibinfo{numpages}{.84}~pages.
\newblock
\showISSN{1084-6654}
\urldef\tempurl%
\url{https://doi.org/10.1145/1498698.1564500}
\showDOI{\tempurl}


\bibitem[\protect\citeauthoryear{Chakrabarti and Wang}{Chakrabarti and
  Wang}{1994}]%
        {ApplicationVideo}
\bibfield{author}{\bibinfo{person}{Chaitali Chakrabarti} {and}
  \bibinfo{person}{Li-Yu Wang}.} \bibinfo{year}{1994}\natexlab{}.
\newblock \showarticletitle{Novel sorting network-based architectures for rank
  order filters}.
\newblock \bibinfo{journal}{\emph{IEEE TVLSI}} \bibinfo{volume}{2},
  \bibinfo{number}{4} (\bibinfo{date}{Dec} \bibinfo{year}{1994}),
  \bibinfo{pages}{502--507}.
\newblock
\showISSN{1063-8210}
\urldef\tempurl%
\url{https://doi.org/10.1109/92.335027}
\showDOI{\tempurl}


\bibitem[\protect\citeauthoryear{Chen and Prasanna}{Chen and Prasanna}{2017}]%
        {Review2Ref3}
\bibfield{author}{\bibinfo{person}{Ren Chen} {and} \bibinfo{person}{Viktor~K.
  Prasanna}.} \bibinfo{year}{2017}\natexlab{}.
\newblock \showarticletitle{Computer Generation of High Throughput and Memory
  Efficient Sorting Designs on FPGA}.
\newblock \bibinfo{journal}{\emph{IEEE Transactions on Parallel and Distributed
  Systems}} \bibinfo{volume}{28}, \bibinfo{number}{11} (\bibinfo{year}{2017}),
  \bibinfo{pages}{3100--3113}.
\newblock
\urldef\tempurl%
\url{https://doi.org/10.1109/TPDS.2017.2705128}
\showDOI{\tempurl}


\bibitem[\protect\citeauthoryear{Cheng, Zheng, Li, Chang, Sze, and Miao}{Cheng
  et~al\mbox{.}}{2020}]%
        {ED2020}
\bibfield{author}{\bibinfo{person}{Long Cheng}, \bibinfo{person}{Hao-Xuan
  Zheng}, \bibinfo{person}{Yi Li}, \bibinfo{person}{Ting-Chang Chang},
  \bibinfo{person}{Simon~M. Sze}, {and} \bibinfo{person}{Xiangshui Miao}.}
  \bibinfo{year}{2020}\natexlab{}.
\newblock \showarticletitle{In-Memory Digital Comparator Based on a Single
  Multivalued One-Transistor-One-Resistor Memristor}.
\newblock \bibinfo{journal}{\emph{IEEE Transactions on Electron Devices}}
  \bibinfo{volume}{67}, \bibinfo{number}{3} (\bibinfo{year}{2020}),
  \bibinfo{pages}{1293--1296}.
\newblock
\urldef\tempurl%
\url{https://doi.org/10.1109/TED.2020.2967401}
\showDOI{\tempurl}


\bibitem[\protect\citeauthoryear{Chu, Luo, Jin, and Wan}{Chu
  et~al\mbox{.}}{2021}]%
        {chunvmsorting}
\bibfield{author}{\bibinfo{person}{Zhaole Chu}, \bibinfo{person}{Yongping Luo},
  \bibinfo{person}{Peiquan Jin}, {and} \bibinfo{person}{Shouhong Wan}.}
  \bibinfo{year}{2021}\natexlab{}.
\newblock \showarticletitle{NVMSorting: Efficient Sorting on Non-Volatile
  Memory}. In \bibinfo{booktitle}{\emph{The 33rd International Conference on
  Software Engineering \& Knowledge Engineering (SEKE 2021)}}.
\newblock


\bibitem[\protect\citeauthoryear{Colavita, Mumolo, and Capello}{Colavita
  et~al\mbox{.}}{1997}]%
        {ApplicationScientific}
\bibfield{author}{\bibinfo{person}{Alberto Colavita}, \bibinfo{person}{Enzo
  Mumolo}, {and} \bibinfo{person}{Gabriele Capello}.}
  \bibinfo{year}{1997}\natexlab{}.
\newblock \showarticletitle{{A novel sorting algorithm and its application to a
  gamma-ray telescope asynchronous data acquisition system}}.
\newblock \bibinfo{journal}{\emph{Nuclear Inst. and Methods in Physics Research
  Sec. A}} \bibinfo{volume}{394}, \bibinfo{number}{3} (\bibinfo{year}{1997}),
  \bibinfo{pages}{374 -- 380}.
\newblock
\showISSN{0168-9002}
\urldef\tempurl%
\url{https://doi.org/10.1016/S0168-9002(97)00567-6}
\showDOI{\tempurl}


\bibitem[\protect\citeauthoryear{Faraji and Bazargan}{Faraji and
  Bazargan}{2020}]%
        {Faraji_Hybrid_TC}
\bibfield{author}{\bibinfo{person}{S.~Rasoul Faraji} {and} \bibinfo{person}{Kia
  Bazargan}.} \bibinfo{year}{2020}\natexlab{}.
\newblock \showarticletitle{Hybrid Binary-Unary Hardware Accelerator}.
\newblock \bibinfo{journal}{\emph{IEEE Trans. Comput.}} \bibinfo{volume}{69},
  \bibinfo{number}{9} (\bibinfo{year}{2020}), \bibinfo{pages}{1308--1319}.
\newblock
\urldef\tempurl%
\url{https://doi.org/10.1109/TC.2020.2971596}
\showDOI{\tempurl}


\bibitem[\protect\citeauthoryear{Farmahini-Farahani, Duwe~III, Schulte, and
  Compton}{Farmahini-Farahani et~al\mbox{.}}{2013}]%
        {SortingPaper1}
\bibfield{author}{\bibinfo{person}{Amin Farmahini-Farahani},
  \bibinfo{person}{Henry~J. Duwe~III}, \bibinfo{person}{Michael~J. Schulte},
  {and} \bibinfo{person}{Katherine Compton}.} \bibinfo{year}{2013}\natexlab{}.
\newblock \showarticletitle{Modular Design of High-Throughput, Low-Latency
  Sorting Units}.
\newblock \bibinfo{journal}{\emph{IEEE Trans. Comput.}} \bibinfo{volume}{62},
  \bibinfo{number}{7} (\bibinfo{year}{2013}), \bibinfo{pages}{1389--1402}.
\newblock
\urldef\tempurl%
\url{https://doi.org/10.1109/TC.2012.108}
\showDOI{\tempurl}


\bibitem[\protect\citeauthoryear{Gaines}{Gaines}{1969}]%
        {Gaines1969}
\bibfield{author}{\bibinfo{person}{Brian~R. Gaines}.}
  \bibinfo{year}{1969}\natexlab{}.
\newblock \showarticletitle{Stochastic Computing Systems}.
\newblock In \bibinfo{booktitle}{\emph{Advances in Information Systems
  Science}}. \bibinfo{publisher}{Springer US}, \bibinfo{pages}{37--172}.
\newblock
\showISBNx{978-1-4899-5843-3}
\urldef\tempurl%
\url{https://doi.org/10.1007/978-1-4899-5841-9_2}
\showDOI{\tempurl}


\bibitem[\protect\citeauthoryear{Gedik, Bordawekar, and Yu}{Gedik
  et~al\mbox{.}}{2007}]%
        {SortingPaper2}
\bibfield{author}{\bibinfo{person}{Bu\v{g}ra Gedik}, \bibinfo{person}{Rajesh~R.
  Bordawekar}, {and} \bibinfo{person}{Philip~S. Yu}.}
  \bibinfo{year}{2007}\natexlab{}.
\newblock \showarticletitle{CellSort: High Performance Sorting on the Cell
  Processor}. In \bibinfo{booktitle}{\emph{VLDB}} (Vienna, Austria).
  \bibinfo{pages}{1286--1297}.
\newblock
\showISBNx{978-1-59593-649-3}


\bibitem[\protect\citeauthoryear{Govindaraju, Gray, Kumar, and
  Manocha}{Govindaraju et~al\mbox{.}}{2006}]%
        {ApplicationDatabase2}
\bibfield{author}{\bibinfo{person}{Naga Govindaraju}, \bibinfo{person}{Jim
  Gray}, \bibinfo{person}{Ritesh Kumar}, {and} \bibinfo{person}{Dinesh
  Manocha}.} \bibinfo{year}{2006}\natexlab{}.
\newblock \showarticletitle{{GPUTeraSort: High Performance Graphics
  Co-processor Sorting for Large Database Management}}. In
  \bibinfo{booktitle}{\emph{SIGMOD}} (Chicago, IL, USA).
  \bibinfo{pages}{325--336}.
\newblock
\showISBNx{1-59593-434-0}
\urldef\tempurl%
\url{https://doi.org/10.1145/1142473.1142511}
\showDOI{\tempurl}


\bibitem[\protect\citeauthoryear{Graefe}{Graefe}{2006}]%
        {ApplicationDatabase}
\bibfield{author}{\bibinfo{person}{Goetz Graefe}.}
  \bibinfo{year}{2006}\natexlab{}.
\newblock \showarticletitle{Implementing Sorting in Database Systems}.
\newblock \bibinfo{journal}{\emph{ACM Comput. Surv.}} \bibinfo{volume}{38},
  \bibinfo{number}{3}, Article \bibinfo{articleno}{10} (\bibinfo{date}{Sept.}
  \bibinfo{year}{2006}).
\newblock
\showISSN{0360-0300}
\urldef\tempurl%
\url{https://doi.org/10.1145/1132960.1132964}
\showDOI{\tempurl}


\bibitem[\protect\citeauthoryear{Gupta, Imani, and Rosing}{Gupta
  et~al\mbox{.}}{2018}]%
        {Felix_imani}
\bibfield{author}{\bibinfo{person}{Saransh Gupta}, \bibinfo{person}{Mohsen
  Imani}, {and} \bibinfo{person}{Tajana Rosing}.}
  \bibinfo{year}{2018}\natexlab{}.
\newblock \showarticletitle{FELIX: Fast and Energy-Efficient Logic in Memory}.
  In \bibinfo{booktitle}{\emph{2018 IEEE/ACM International Conference on
  Computer-Aided Design (ICCAD)}}. \bibinfo{pages}{1--7}.
\newblock
\urldef\tempurl%
\url{https://doi.org/10.1145/3240765.3240811}
\showDOI{\tempurl}


\bibitem[\protect\citeauthoryear{Gupta, Imani, Sim, Huang, Wu, Najafi, and
  Rosing}{Gupta et~al\mbox{.}}{2020}]%
        {Scrimp_Gupta20}
\bibfield{author}{\bibinfo{person}{Saransh Gupta}, \bibinfo{person}{Mohsen
  Imani}, \bibinfo{person}{Joonseop Sim}, \bibinfo{person}{Andrew Huang},
  \bibinfo{person}{Fan Wu}, \bibinfo{person}{M.~Hassan Najafi}, {and}
  \bibinfo{person}{Tajana Rosing}.} \bibinfo{year}{2020}\natexlab{}.
\newblock \showarticletitle{SCRIMP: A General Stochastic Computing Architecture
  using ReRAM in-Memory Processing}. In \bibinfo{booktitle}{\emph{2020 Design,
  Automation Test in Europe Conference Exhibition (DATE)}}.
  \bibinfo{pages}{1598--1601}.
\newblock
\urldef\tempurl%
\url{https://doi.org/10.23919/DATE48585.2020.9116338}
\showDOI{\tempurl}


\bibitem[\protect\citeauthoryear{Hamdioui, Kvatinsky, Cauwenberghs, Xie, Wald,
  Joshi, Elsayed, Corporaal, and Bertels}{Hamdioui et~al\mbox{.}}{2017}]%
        {Hamdioui2017DATE}
\bibfield{author}{\bibinfo{person}{Said Hamdioui}, \bibinfo{person}{Shahar
  Kvatinsky}, \bibinfo{person}{Gert Cauwenberghs}, \bibinfo{person}{Lei Xie},
  \bibinfo{person}{Nimrod Wald}, \bibinfo{person}{Siddharth Joshi},
  \bibinfo{person}{Hesham~Mostafa Elsayed}, \bibinfo{person}{Henk Corporaal},
  {and} \bibinfo{person}{Koen Bertels}.} \bibinfo{year}{2017}\natexlab{}.
\newblock \showarticletitle{Memristor for Computing: Myth or Reality?}
  \emph{(\bibinfo{series}{DATE ’17})}.
\newblock


\bibitem[\protect\citeauthoryear{Jalilvand, Najafi, and Fazeli}{Jalilvand
  et~al\mbox{.}}{2020}]%
        {ISCAS20_Jalilvand_Fuzzy}
\bibfield{author}{\bibinfo{person}{Amir~Hossein Jalilvand},
  \bibinfo{person}{M.~Hassan Najafi}, {and} \bibinfo{person}{Mahdi Fazeli}.}
  \bibinfo{year}{2020}\natexlab{}.
\newblock \showarticletitle{Fuzzy-logic using Unary Bit-Stream Processing}. In
  \bibinfo{booktitle}{\emph{2020 IEEE International Symposium on Circuits and
  Systems (ISCAS)}}. \bibinfo{pages}{1--5}.
\newblock


\bibitem[\protect\citeauthoryear{Jeloka, Akesh, Sylvester, and Blaauw}{Jeloka
  et~al\mbox{.}}{2016}]%
        {EqualityComparator2}
\bibfield{author}{\bibinfo{person}{Supreet Jeloka},
  \bibinfo{person}{Naveen~Bharathwaj Akesh}, \bibinfo{person}{Dennis
  Sylvester}, {and} \bibinfo{person}{David Blaauw}.}
  \bibinfo{year}{2016}\natexlab{}.
\newblock \showarticletitle{A 28 nm Configurable Memory (TCAM/BCAM/SRAM) Using
  Push-Rule 6T Bit Cell Enabling Logic-in-Memory}.
\newblock \bibinfo{journal}{\emph{IEEE Journal of Solid-State Circuits}}
  \bibinfo{volume}{51}, \bibinfo{number}{4} (\bibinfo{year}{2016}),
  \bibinfo{pages}{1009--1021}.
\newblock
\urldef\tempurl%
\url{https://doi.org/10.1109/JSSC.2016.2515510}
\showDOI{\tempurl}


\bibitem[\protect\citeauthoryear{Keckler, Dally, Khailany, Garland, and
  Glasco}{Keckler et~al\mbox{.}}{2011}]%
        {memory_energy_wire}
\bibfield{author}{\bibinfo{person}{Stephen~W. Keckler},
  \bibinfo{person}{William~J. Dally}, \bibinfo{person}{Brucek Khailany},
  \bibinfo{person}{Michael Garland}, {and} \bibinfo{person}{David Glasco}.}
  \bibinfo{year}{2011}\natexlab{}.
\newblock \showarticletitle{GPUs and the Future of Parallel Computing}.
\newblock \bibinfo{journal}{\emph{IEEE Micro}} \bibinfo{volume}{31},
  \bibinfo{number}{5} (\bibinfo{year}{2011}), \bibinfo{pages}{7--17}.
\newblock
\urldef\tempurl%
\url{https://doi.org/10.1109/MM.2011.89}
\showDOI{\tempurl}


\bibitem[\protect\citeauthoryear{Koch and Torresen}{Koch and Torresen}{2011}]%
        {Sorting_FPGA1}
\bibfield{author}{\bibinfo{person}{Dirk Koch} {and} \bibinfo{person}{Jim
  Torresen}.} \bibinfo{year}{2011}\natexlab{}.
\newblock \showarticletitle{{FPGASort: A High Performance Sorting Architecture
  Exploiting Run-time Reconfiguration on FPGAs for Large Problem Sorting}}. In
  \bibinfo{booktitle}{\emph{FPGA}}.
\newblock
\showISBNx{978-1-4503-0554-9}
\urldef\tempurl%
\url{https://doi.org/10.1145/1950413.1950427}
\showDOI{\tempurl}


\bibitem[\protect\citeauthoryear{Koo, Matam, I., Narra, Li, Tseng, Swanson, and
  Annavaram}{Koo et~al\mbox{.}}{2017}]%
        {Koo_Micro17}
\bibfield{author}{\bibinfo{person}{Gunjae Koo}, \bibinfo{person}{Kiran~Kumar
  Matam}, \bibinfo{person}{Te I.}, \bibinfo{person}{H.V. Krishna~Giri Narra},
  \bibinfo{person}{Jing Li}, \bibinfo{person}{Hung-Wei Tseng},
  \bibinfo{person}{Steven Swanson}, {and} \bibinfo{person}{Murali Annavaram}.}
  \bibinfo{year}{2017}\natexlab{}.
\newblock \showarticletitle{Summarizer: Trading Communication with Computing
  Near Storage}. In \bibinfo{booktitle}{\emph{2017 50th Annual IEEE/ACM
  International Symposium on Microarchitecture (MICRO)}}.
  \bibinfo{pages}{219--231}.
\newblock


\bibitem[\protect\citeauthoryear{Kvatinsky, Belousov, Liman, Satat, Wald,
  Friedman, Kolodny, and Weiser}{Kvatinsky et~al\mbox{.}}{2014}]%
        {MAGIC_2014}
\bibfield{author}{\bibinfo{person}{Shahar Kvatinsky}, \bibinfo{person}{Dmitry
  Belousov}, \bibinfo{person}{Slavik Liman}, \bibinfo{person}{Guy Satat},
  \bibinfo{person}{Nimrod Wald}, \bibinfo{person}{Eby~G. Friedman},
  \bibinfo{person}{Avinoam Kolodny}, {and} \bibinfo{person}{Uri~C. Weiser}.}
  \bibinfo{year}{2014}\natexlab{}.
\newblock \showarticletitle{MAGIC—Memristor-Aided Logic}.
\newblock \bibinfo{journal}{\emph{IEEE Transactions on Circuits and Systems II:
  Express Briefs}} \bibinfo{volume}{61}, \bibinfo{number}{11}
  (\bibinfo{year}{2014}), \bibinfo{pages}{895--899}.
\newblock
\urldef\tempurl%
\url{https://doi.org/10.1109/TCSII.2014.2357292}
\showDOI{\tempurl}


\bibitem[\protect\citeauthoryear{Kvatinsky, Ramadan, Friedman, and
  Kolodny}{Kvatinsky et~al\mbox{.}}{2015}]%
        {VTEAM}
\bibfield{author}{\bibinfo{person}{Shahar Kvatinsky}, \bibinfo{person}{Misbah
  Ramadan}, \bibinfo{person}{Eby~G. Friedman}, {and} \bibinfo{person}{Avinoam
  Kolodny}.} \bibinfo{year}{2015}\natexlab{}.
\newblock \showarticletitle{VTEAM: A General Model for Voltage-Controlled
  Memristors}.
\newblock \bibinfo{journal}{\emph{IEEE Transactions on Circuits and Systems II:
  Express Briefs}} \bibinfo{volume}{62}, \bibinfo{number}{8}
  (\bibinfo{year}{2015}), \bibinfo{pages}{786--790}.
\newblock
\urldef\tempurl%
\url{https://doi.org/10.1109/TCSII.2015.2433536}
\showDOI{\tempurl}


\bibitem[\protect\citeauthoryear{Li, Lilja, Qian, Bazargan, and Riedel}{Li
  et~al\mbox{.}}{2014}]%
        {Peng_TVLSI14}
\bibfield{author}{\bibinfo{person}{Peng Li}, \bibinfo{person}{D.J. Lilja},
  \bibinfo{person}{Weikang Qian}, \bibinfo{person}{K. Bazargan}, {and}
  \bibinfo{person}{M.D. Riedel}.} \bibinfo{year}{2014}\natexlab{}.
\newblock \showarticletitle{Computation on Stochastic Bit Streams Digital Image
  Processing Case Studies}.
\newblock \bibinfo{journal}{\emph{IEEE TVLSI}} \bibinfo{volume}{22},
  \bibinfo{number}{3} (\bibinfo{year}{2014}), \bibinfo{pages}{449--462}.
\newblock
\showISSN{1063-8210}
\urldef\tempurl%
\url{https://doi.org/10.1109/TVLSI.2013.2247429}
\showDOI{\tempurl}


\bibitem[\protect\citeauthoryear{Li, Challapalle, Ramanathan, and Narayanan}{Li
  et~al\mbox{.}}{2020}]%
        {Li_GLSVLSI20}
\bibfield{author}{\bibinfo{person}{Zheyu Li}, \bibinfo{person}{Nagadastagiri
  Challapalle}, \bibinfo{person}{Akshay~Krishna Ramanathan}, {and}
  \bibinfo{person}{Vijaykrishnan Narayanan}.} \bibinfo{year}{2020}\natexlab{}.
\newblock \showarticletitle{IMC-Sort: In-Memory Parallel Sorting Architecture
  Using Hybrid Memory Cube}. In \bibinfo{booktitle}{\emph{Proceedings of the
  2020 on Great Lakes Symposium on VLSI}} (Virtual Event, China)
  \emph{(\bibinfo{series}{GLSVLSI '20})}. \bibinfo{publisher}{Association for
  Computing Machinery}, \bibinfo{address}{New York, NY, USA},
  \bibinfo{pages}{45–50}.
\newblock
\showISBNx{9781450379441}
\urldef\tempurl%
\url{https://doi.org/10.1145/3386263.3407581}
\showDOI{\tempurl}


\bibitem[\protect\citeauthoryear{Madhavan, Sherwood, and Strukov}{Madhavan
  et~al\mbox{.}}{2014}]%
        {RaceLogic}
\bibfield{author}{\bibinfo{person}{Advait Madhavan}, \bibinfo{person}{Timothy
  Sherwood}, {and} \bibinfo{person}{Dmitri Strukov}.}
  \bibinfo{year}{2014}\natexlab{}.
\newblock \showarticletitle{Race Logic: A hardware acceleration for dynamic
  programming algorithms}. In \bibinfo{booktitle}{\emph{2014 ACM/IEEE 41st
  International Symposium on Computer Architecture (ISCA)}}.
  \bibinfo{pages}{517--528}.
\newblock
\urldef\tempurl%
\url{https://doi.org/10.1109/ISCA.2014.6853226}
\showDOI{\tempurl}


\bibitem[\protect\citeauthoryear{Mohajer, Wang, and Bazargan}{Mohajer
  et~al\mbox{.}}{2018}]%
        {RoutingMagic_FPGA_2018}
\bibfield{author}{\bibinfo{person}{Soheil Mohajer}, \bibinfo{person}{Zhiheng
  Wang}, {and} \bibinfo{person}{Kia Bazargan}.}
  \bibinfo{year}{2018}\natexlab{}.
\newblock \showarticletitle{Routing Magic: Performing Computations Using
  Routing Networks and Voting Logic on Unary Encoded Data}. In
  \bibinfo{booktitle}{\emph{Proceedings of the 2018 ACM/SIGDA International
  Symposium on Field-Programmable Gate Arrays}} (Monterey, CALIFORNIA, USA)
  \emph{(\bibinfo{series}{FPGA '18})}. \bibinfo{publisher}{Association for
  Computing Machinery}, \bibinfo{address}{New York, NY, USA},
  \bibinfo{pages}{77–86}.
\newblock
\showISBNx{9781450356145}
\urldef\tempurl%
\url{https://doi.org/10.1145/3174243.3174267}
\showDOI{\tempurl}


\bibitem[\protect\citeauthoryear{{Najafi}, {Faraji}, {Bazargan}, and
  {Lilja}}{{Najafi} et~al\mbox{.}}{2020}]%
        {Najafi_Pulse_ISCAS20}
\bibfield{author}{\bibinfo{person}{M.~Hassan {Najafi}},
  \bibinfo{person}{S.~Rasoul {Faraji}}, \bibinfo{person}{Kia {Bazargan}}, {and}
  \bibinfo{person}{David {Lilja}}.} \bibinfo{year}{2020}\natexlab{}.
\newblock \showarticletitle{Energy-Efficient Pulse-Based Convolution for
  Near-Sensor Processing}. In \bibinfo{booktitle}{\emph{2020 IEEE International
  Symposium on Circuits and Systems (ISCAS)}}. \bibinfo{pages}{1--5}.
\newblock
\urldef\tempurl%
\url{https://doi.org/10.1109/ISCAS45731.2020.9181248}
\showDOI{\tempurl}


\bibitem[\protect\citeauthoryear{Najafi, Lilja, Riedel, and Bazargan}{Najafi
  et~al\mbox{.}}{2018}]%
        {Sorting-TVLSI-2018}
\bibfield{author}{\bibinfo{person}{M.~Hassan Najafi}, \bibinfo{person}{David
  Lilja}, \bibinfo{person}{Marc~D. Riedel}, {and} \bibinfo{person}{Kia
  Bazargan}.} \bibinfo{year}{2018}\natexlab{}.
\newblock \showarticletitle{{Low-Cost Sorting Network Circuits Using Unary
  Processing}}.
\newblock \bibinfo{journal}{\emph{IEEE Trans. on VLSI Systems}}
  \bibinfo{volume}{26}, \bibinfo{number}{8} (\bibinfo{date}{Aug}
  \bibinfo{year}{2018}), \bibinfo{pages}{1471--1480}.
\newblock
\showISSN{1063-8210}
\urldef\tempurl%
\url{https://doi.org/10.1109/TVLSI.2018.2822300}
\showDOI{\tempurl}


\bibitem[\protect\citeauthoryear{{Najafi}, {Lilja}, {Riedel}, and
  {Bazargan}}{{Najafi} et~al\mbox{.}}{2017}]%
        {Najafi_Sorting_ICCD2017}
\bibfield{author}{\bibinfo{person}{M.~Hassan {Najafi}},
  \bibinfo{person}{David~J. {Lilja}}, \bibinfo{person}{Marc {Riedel}}, {and}
  \bibinfo{person}{Kia {Bazargan}}.} \bibinfo{year}{2017}\natexlab{}.
\newblock \showarticletitle{Power and Area Efficient Sorting Networks Using
  Unary Processing}. In \bibinfo{booktitle}{\emph{2017 IEEE International
  Conference on Computer Design (ICCD)}}. \bibinfo{pages}{125--128}.
\newblock
\urldef\tempurl%
\url{https://doi.org/10.1109/ICCD.2017.27}
\showDOI{\tempurl}


\bibitem[\protect\citeauthoryear{Nguyen, Yu, Lebdeh, Taouil, Hamdioui, and
  Catthoor}{Nguyen et~al\mbox{.}}{2020}]%
        {Class20}
\bibfield{author}{\bibinfo{person}{Hoang Anh~Du Nguyen},
  \bibinfo{person}{Jintao Yu}, \bibinfo{person}{Muath~Abu Lebdeh},
  \bibinfo{person}{Mottaqiallah Taouil}, \bibinfo{person}{Said Hamdioui}, {and}
  \bibinfo{person}{Francky Catthoor}.} \bibinfo{year}{2020}\natexlab{}.
\newblock \showarticletitle{A Classification of Memory-Centric Computing}.
\newblock \bibinfo{journal}{\emph{J. Emerg. Technol. Comput. Syst.}}
  \bibinfo{volume}{16}, \bibinfo{number}{2}, Article \bibinfo{articleno}{13}
  (\bibinfo{date}{Jan.} \bibinfo{year}{2020}), \bibinfo{numpages}{26}~pages.
\newblock
\showISSN{1550-4832}
\urldef\tempurl%
\url{https://doi.org/10.1145/3365837}
\showDOI{\tempurl}


\bibitem[\protect\citeauthoryear{Nikahd, Behnam, and Sameni}{Nikahd
  et~al\mbox{.}}{2016}]%
        {MedianPaper3}
\bibfield{author}{\bibinfo{person}{Eesa Nikahd}, \bibinfo{person}{Payman
  Behnam}, {and} \bibinfo{person}{Reza Sameni}.}
  \bibinfo{year}{2016}\natexlab{}.
\newblock \showarticletitle{High-Speed Hardware Implementation of Fixed and
  Runtime Variable Window Length 1-D Median Filters}.
\newblock \bibinfo{journal}{\emph{IEEE Tran. on Circ. \& Sys. II: Express
  Briefs}} (\bibinfo{year}{2016}).
\newblock
\showISSN{1549-7747}
\urldef\tempurl%
\url{https://doi.org/10.1109/TCSII.2015.2504945}
\showDOI{\tempurl}


\bibitem[\protect\citeauthoryear{{Olarlu}, {Pinotti}, and {Si Qing
  Zheng}}{{Olarlu} et~al\mbox{.}}{2000}]%
        {Sorting_ASIC}
\bibfield{author}{\bibinfo{person}{Stephan {Olarlu}},
  \bibinfo{person}{M.~Cristina {Pinotti}}, {and} \bibinfo{person}{{Si Qing
  Zheng}}.} \bibinfo{year}{2000}\natexlab{}.
\newblock \showarticletitle{An optimal hardware-algorithm for sorting using a
  fixed-size parallel sorting device}.
\newblock \bibinfo{journal}{\emph{IEEE Trans. Comput.}} \bibinfo{volume}{49},
  \bibinfo{number}{12} (\bibinfo{year}{2000}), \bibinfo{pages}{1310--1324}.
\newblock
\urldef\tempurl%
\url{https://doi.org/10.1109/12.895849}
\showDOI{\tempurl}


\bibitem[\protect\citeauthoryear{Pagiamtzis and Sheikholeslami}{Pagiamtzis and
  Sheikholeslami}{2006}]%
        {CAM2006}
\bibfield{author}{\bibinfo{person}{Kostas Pagiamtzis} {and}
  \bibinfo{person}{Ali Sheikholeslami}.} \bibinfo{year}{2006}\natexlab{}.
\newblock \showarticletitle{Content-addressable memory (CAM) circuits and
  architectures: a tutorial and survey}.
\newblock \bibinfo{journal}{\emph{IEEE Journal of Solid-State Circuits}}
  \bibinfo{volume}{41}, \bibinfo{number}{3} (\bibinfo{year}{2006}),
  \bibinfo{pages}{712--727}.
\newblock
\urldef\tempurl%
\url{https://doi.org/10.1109/JSSC.2005.864128}
\showDOI{\tempurl}


\bibitem[\protect\citeauthoryear{Pok, Chen, Schamus, Montgomery, and Tsui}{Pok
  et~al\mbox{.}}{1997}]%
        {ApplicationSignal}
\bibfield{author}{\bibinfo{person}{David~K. Pok}, \bibinfo{person}{Chien-In
  Chen}, \bibinfo{person}{John~J. Schamus}, \bibinfo{person}{Christine~T.
  Montgomery}, {and} \bibinfo{person}{James B.~Y. Tsui}.}
  \bibinfo{year}{1997}\natexlab{}.
\newblock \showarticletitle{Chip design for monobit receiver}.
\newblock \bibinfo{journal}{\emph{IEEE Trans. on Microwave Theory and
  Techniques}} \bibinfo{volume}{45}, \bibinfo{number}{12} (\bibinfo{date}{Dec}
  \bibinfo{year}{1997}), \bibinfo{pages}{2283--2295}.
\newblock
\showISSN{0018-9480}
\urldef\tempurl%
\url{https://doi.org/10.1109/22.643832}
\showDOI{\tempurl}


\bibitem[\protect\citeauthoryear{Poppelbaum}{Poppelbaum}{1978}]%
        {PoppelbaumUnary2}
\bibfield{author}{\bibinfo{person}{Wolfgang~J. Poppelbaum}.}
  \bibinfo{year}{1978}\natexlab{}.
\newblock \showarticletitle{Burst Processing: A Deterministic Counterpart to
  Stochastic Computing}.
\newblock In \bibinfo{booktitle}{\emph{Proceedings of the 1st Intern. Symp. on
  Stochastic Computing and its Apps.}}
\newblock


\bibitem[\protect\citeauthoryear{Poppelbaum, Dollas, Glickman, and
  O'Toole}{Poppelbaum et~al\mbox{.}}{1987}]%
        {Poppelbaum198747}
\bibfield{author}{\bibinfo{person}{Wolfgang~J. Poppelbaum}, \bibinfo{person}{A.
  Dollas}, \bibinfo{person}{J.B. Glickman}, {and} \bibinfo{person}{C.
  O'Toole}.} \bibinfo{year}{1987}\natexlab{}.
\newblock \showarticletitle{Unary Processing}.
\newblock In \bibinfo{booktitle}{\emph{Advances in Computers}}.
  Vol.~\bibinfo{volume}{26}. \bibinfo{publisher}{Elsevier}, \bibinfo{pages}{47
  -- 92}.
\newblock
\showISSN{0065-2458}
\urldef\tempurl%
\url{https://doi.org/10.1016/S0065-2458(08)60005-4}
\showDOI{\tempurl}


\bibitem[\protect\citeauthoryear{Prasad, Rezaalipour, Dehyadegari, and
  Bojnordi}{Prasad et~al\mbox{.}}{2021}]%
        {Prasad_HPCA21}
\bibfield{author}{\bibinfo{person}{Ananth~Krishna Prasad},
  \bibinfo{person}{Morteza Rezaalipour}, \bibinfo{person}{Masoud Dehyadegari},
  {and} \bibinfo{person}{Mahdi~Nazm Bojnordi}.}
  \bibinfo{year}{2021}\natexlab{}.
\newblock \showarticletitle{Memristive Data Ranking}. In
  \bibinfo{booktitle}{\emph{2021 IEEE International Symposium on
  High-Performance Computer Architecture (HPCA)}}. \bibinfo{pages}{440--452}.
\newblock
\urldef\tempurl%
\url{https://doi.org/10.1109/HPCA51647.2021.00045}
\showDOI{\tempurl}


\bibitem[\protect\citeauthoryear{Pugsley, Deb, Balasubramonian, and Li}{Pugsley
  et~al\mbox{.}}{2015}]%
        {ICCD2015}
\bibfield{author}{\bibinfo{person}{Seth~H. Pugsley}, \bibinfo{person}{Arjun
  Deb}, \bibinfo{person}{Rajeev Balasubramonian}, {and} \bibinfo{person}{Feifei
  Li}.} \bibinfo{year}{2015}\natexlab{}.
\newblock \showarticletitle{Fixed-function hardware sorting accelerators for
  near data MapReduce execution}. In \bibinfo{booktitle}{\emph{2015 33rd IEEE
  International Conference on Computer Design (ICCD)}}.
  \bibinfo{pages}{439--442}.
\newblock
\urldef\tempurl%
\url{https://doi.org/10.1109/ICCD.2015.7357143}
\showDOI{\tempurl}


\bibitem[\protect\citeauthoryear{Qiao, Oh, Guo, Chang, and Cong}{Qiao
  et~al\mbox{.}}{2021}]%
        {Qiao21}
\bibfield{author}{\bibinfo{person}{Weikang Qiao}, \bibinfo{person}{Jihun Oh},
  \bibinfo{person}{Licheng Guo}, \bibinfo{person}{Mau-Chung~Frank Chang}, {and}
  \bibinfo{person}{Jason Cong}.} \bibinfo{year}{2021}\natexlab{}.
\newblock \showarticletitle{FANS: FPGA-Accelerated Near-Storage Sorting}. In
  \bibinfo{booktitle}{\emph{2021 IEEE 29th Annual International Symposium on
  Field-Programmable Custom Computing Machines (FCCM)}}.
  \bibinfo{pages}{106--114}.
\newblock
\urldef\tempurl%
\url{https://doi.org/10.1109/FCCM51124.2021.00020}
\showDOI{\tempurl}


\bibitem[\protect\citeauthoryear{Radakovits and TaheriNejad}{Radakovits and
  TaheriNejad}{2019}]%
        {Radakovits2019}
\bibfield{author}{\bibinfo{person}{David Radakovits} {and}
  \bibinfo{person}{Nima TaheriNejad}.} \bibinfo{year}{2019}\natexlab{}.
\newblock \showarticletitle{Implementation and Characterization of a Memristive
  Memory System}. In \bibinfo{booktitle}{\emph{2019 IEEE 32nd Canadian
  Conference on Electrical and Computer Engineering (CCECE)}}.
  \bibinfo{pages}{1--5}.
\newblock


\bibitem[\protect\citeauthoryear{Ruan, He, and Cong}{Ruan
  et~al\mbox{.}}{2019}]%
        {Ruan_ICCAD19}
\bibfield{author}{\bibinfo{person}{Zhenyuan Ruan}, \bibinfo{person}{Tong He},
  {and} \bibinfo{person}{Jason Cong}.} \bibinfo{year}{2019}\natexlab{}.
\newblock \showarticletitle{Analyzing and Modeling In-Storage Computing
  Workloads On EISC — An FPGA-Based System-Level Emulation Platform}. In
  \bibinfo{booktitle}{\emph{2019 IEEE/ACM International Conference on
  Computer-Aided Design (ICCAD)}}. \bibinfo{pages}{1--8}.
\newblock
\urldef\tempurl%
\url{https://doi.org/10.1109/ICCAD45719.2019.8942135}
\showDOI{\tempurl}


\bibitem[\protect\citeauthoryear{Salamat, Haj~Aboutalebi, Khaleghi, Lee, Ki,
  and Rosing}{Salamat et~al\mbox{.}}{2021}]%
        {salamat2021nascent}
\bibfield{author}{\bibinfo{person}{Sahand Salamat}, \bibinfo{person}{Armin
  Haj~Aboutalebi}, \bibinfo{person}{Behnam Khaleghi}, \bibinfo{person}{Joo~Hwan
  Lee}, \bibinfo{person}{Yang~Seok Ki}, {and} \bibinfo{person}{Tajana Rosing}.}
  \bibinfo{year}{2021}\natexlab{}.
\newblock \showarticletitle{NASCENT: Near-Storage Acceleration of Database Sort
  on SmartSSD}. In \bibinfo{booktitle}{\emph{The 2021 ACM/SIGDA International
  Symposium on Field-Programmable Gate Arrays}}. \bibinfo{pages}{262--272}.
\newblock


\bibitem[\protect\citeauthoryear{Samardzic, Qiao, Aggarwal, Chang, and
  Cong}{Samardzic et~al\mbox{.}}{2020}]%
        {bonsai2020}
\bibfield{author}{\bibinfo{person}{Nikola Samardzic}, \bibinfo{person}{Weikang
  Qiao}, \bibinfo{person}{Vaibhav Aggarwal}, \bibinfo{person}{Mau-Chung~Frank
  Chang}, {and} \bibinfo{person}{Jason Cong}.} \bibinfo{year}{2020}\natexlab{}.
\newblock \showarticletitle{Bonsai: High-Performance Adaptive Merge Tree
  Sorting}. In \bibinfo{booktitle}{\emph{2020 ACM/IEEE 47th Annual
  International Symposium on Computer Architecture (ISCA)}}.
  \bibinfo{pages}{282--294}.
\newblock
\urldef\tempurl%
\url{https://doi.org/10.1109/ISCA45697.2020.00033}
\showDOI{\tempurl}


\bibitem[\protect\citeauthoryear{Satish, Harris, and Garland}{Satish
  et~al\mbox{.}}{2009}]%
        {Sorting_GPU2}
\bibfield{author}{\bibinfo{person}{Nadathur Satish}, \bibinfo{person}{Mark
  Harris}, {and} \bibinfo{person}{Michael Garland}.}
  \bibinfo{year}{2009}\natexlab{}.
\newblock \showarticletitle{Designing efficient sorting algorithms for manycore
  GPUs}. In \bibinfo{booktitle}{\emph{2009 IEEE International Symposium on
  Parallel Distributed Processing}}. \bibinfo{pages}{1--10}.
\newblock
\urldef\tempurl%
\url{https://doi.org/10.1109/IPDPS.2009.5161005}
\showDOI{\tempurl}


\bibitem[\protect\citeauthoryear{Smith}{Smith}{2018}]%
        {Smith_ISCA_2018}
\bibfield{author}{\bibinfo{person}{James~E. Smith}.}
  \bibinfo{year}{2018}\natexlab{}.
\newblock \showarticletitle{Space-time Algebra: A Model for Neocortical
  Computation}. In \bibinfo{booktitle}{\emph{ISCA'18}} (Los Angeles,
  California). \bibinfo{pages}{289--300}.
\newblock
\showISBNx{978-1-5386-5984-7}
\urldef\tempurl%
\url{https://doi.org/10.1109/ISCA.2018.00033}
\showDOI{\tempurl}


\bibitem[\protect\citeauthoryear{Stephens, Bennett, and Zhang}{Stephens
  et~al\mbox{.}}{1999}]%
        {ApplicationScheduling}
\bibfield{author}{\bibinfo{person}{Donpaul~C. Stephens}, \bibinfo{person}{Jon
  C.~R. Bennett}, {and} \bibinfo{person}{Hui Zhang}.}
  \bibinfo{year}{1999}\natexlab{}.
\newblock \showarticletitle{Implementing scheduling algorithms in high-speed
  networks}.
\newblock \bibinfo{journal}{\emph{IEEE JSAC}} \bibinfo{volume}{17},
  \bibinfo{number}{6} (\bibinfo{date}{Jun} \bibinfo{year}{1999}),
  \bibinfo{pages}{1145--1158}.
\newblock
\showISSN{0733-8716}
\urldef\tempurl%
\url{https://doi.org/10.1109/49.772449}
\showDOI{\tempurl}


\bibitem[\protect\citeauthoryear{TaheriNejad}{TaheriNejad}{2021}]%
        {Taherinejad2021tvlsi}
\bibfield{author}{\bibinfo{person}{Nima TaheriNejad}.}
  \bibinfo{year}{2021}\natexlab{}.
\newblock \showarticletitle{{SIXOR:} Single-cycle In-memristor {XOR}}.
\newblock \bibinfo{journal}{\emph{IEEE Transactions on Very Large Scale
  Integration Systems {(TVLSI)}}} \bibinfo{volume}{29}, \bibinfo{number}{5}
  (\bibinfo{year}{2021}), \bibinfo{pages}{925--935}.
\newblock
\urldef\tempurl%
\url{https://doi.org/10.1109/TVLSI.2021.3062293}
\showDOI{\tempurl}


\bibitem[\protect\citeauthoryear{Taherinejad, Manoj, and Jantsch}{Taherinejad
  et~al\mbox{.}}{2015}]%
        {Taherinejad2015}
\bibfield{author}{\bibinfo{person}{Nima Taherinejad},
  \bibinfo{person}{P.~D.~Sai Manoj}, {and} \bibinfo{person}{Axel Jantsch}.}
  \bibinfo{year}{2015}\natexlab{}.
\newblock \showarticletitle{Memristors' Potential for Multi-bit Storage and
  Pattern Learning}. In \bibinfo{booktitle}{\emph{2015 IEEE European Modelling
  Symposium (EMS)}}. \bibinfo{pages}{450--455}.
\newblock
\urldef\tempurl%
\url{https://doi.org/10.1109/EMS.2015.73}
\showDOI{\tempurl}


\bibitem[\protect\citeauthoryear{TaheriNejad and Radakovits}{TaheriNejad and
  Radakovits}{2019}]%
        {Taherinejad2019cas}
\bibfield{author}{\bibinfo{person}{N. TaheriNejad} {and} \bibinfo{person}{D.
  Radakovits}.} \bibinfo{year}{2019}\natexlab{}.
\newblock \showarticletitle{From Behavioral Design of Memristive Circuits and
  Systems to Physical Implementations}.
\newblock \bibinfo{journal}{\emph{IEEE Circuit and Systems (CAS) Magazine}}
  (\bibinfo{year}{2019}), \bibinfo{pages}{1--11}.
\newblock


\bibitem[\protect\citeauthoryear{Taherinejad, Sai, Rathmair, and
  Jantsch}{Taherinejad et~al\mbox{.}}{2016}]%
        {Taherinejad2016}
\bibfield{author}{\bibinfo{person}{Nima Taherinejad},
  \bibinfo{person}{Manoj~P.D. Sai}, \bibinfo{person}{Michael Rathmair}, {and}
  \bibinfo{person}{Axel Jantsch}.} \bibinfo{year}{2016}\natexlab{}.
\newblock \showarticletitle{Fully digital write-in scheme for multi-bit
  memristive storage}. In \bibinfo{booktitle}{\emph{2016 13th International
  Conference on Electrical Engineering, Computing Science and Automatic Control
  (CCE)}}. \bibinfo{pages}{1--6}.
\newblock
\urldef\tempurl%
\url{https://doi.org/10.1109/ICEEE.2016.7751193}
\showDOI{\tempurl}


\bibitem[\protect\citeauthoryear{Tseng, Liu, Gahagan, Li, Jin, and
  Swanson}{Tseng et~al\mbox{.}}{2015}]%
        {Tseng2015GullfossA}
\bibfield{author}{\bibinfo{person}{Hung-Wei Tseng}, \bibinfo{person}{Yang Liu},
  \bibinfo{person}{Mark Gahagan}, \bibinfo{person}{Jing Li},
  \bibinfo{person}{Yanqin Jin}, {and} \bibinfo{person}{Steven Swanson}.}
  \bibinfo{year}{2015}\natexlab{}.
\newblock \showarticletitle{Gullfoss : Accelerating and Simplifying Data
  Movement among Heterogeneous Computing and Storage Resources}.
\newblock


\bibitem[\protect\citeauthoryear{Tzimpragos, Vasudevan, Tsiskaridze,
  Michelogiannakis, Madhavan, Volk, Shalf, and Sherwood}{Tzimpragos
  et~al\mbox{.}}{2020}]%
        {Temporal_ASPLOS20}
\bibfield{author}{\bibinfo{person}{Georgios Tzimpragos}, \bibinfo{person}{Dilip
  Vasudevan}, \bibinfo{person}{Nestan Tsiskaridze}, \bibinfo{person}{George
  Michelogiannakis}, \bibinfo{person}{Advait Madhavan},
  \bibinfo{person}{Jennifer Volk}, \bibinfo{person}{John Shalf}, {and}
  \bibinfo{person}{Timothy Sherwood}.} \bibinfo{year}{2020}\natexlab{}.
\newblock \showarticletitle{A Computational Temporal Logic for Superconducting
  Accelerators}. In \bibinfo{booktitle}{\emph{Proceedings of the Twenty-Fifth
  International Conference on Architectural Support for Programming Languages
  and Operating Systems}} (Lausanne, Switzerland)
  \emph{(\bibinfo{series}{ASPLOS '20})}. \bibinfo{publisher}{Association for
  Computing Machinery}, \bibinfo{address}{New York, NY, USA},
  \bibinfo{pages}{435–448}.
\newblock
\showISBNx{9781450371025}
\urldef\tempurl%
\url{https://doi.org/10.1145/3373376.3378517}
\showDOI{\tempurl}


\bibitem[\protect\citeauthoryear{Wu, Li, Yin, Hsiao, Kim, and Miguel}{Wu
  et~al\mbox{.}}{2020}]%
        {UGEMM2020}
\bibfield{author}{\bibinfo{person}{Di Wu}, \bibinfo{person}{Jingjie Li},
  \bibinfo{person}{Ruokai Yin}, \bibinfo{person}{Hsuan Hsiao},
  \bibinfo{person}{Younghyun Kim}, {and} \bibinfo{person}{Joshua~San Miguel}.}
  \bibinfo{year}{2020}\natexlab{}.
\newblock \showarticletitle{UGEMM: Unary Computing Architecture for GEMM
  Applications}. In \bibinfo{booktitle}{\emph{2020 ACM/IEEE 47th Annual
  International Symposium on Computer Architecture (ISCA)}}.
  \bibinfo{pages}{377--390}.
\newblock
\urldef\tempurl%
\url{https://doi.org/10.1109/ISCA45697.2020.00040}
\showDOI{\tempurl}


\bibitem[\protect\citeauthoryear{Xie, Du~Nguyen, Yu, Kaichouhi, Taouil,
  AlFailakawi, and Hamdioui}{Xie et~al\mbox{.}}{2017}]%
        {xie2017scouting}
\bibfield{author}{\bibinfo{person}{Lei Xie}, \bibinfo{person}{Hoang~Anh
  Du~Nguyen}, \bibinfo{person}{Jintao Yu}, \bibinfo{person}{Ali Kaichouhi},
  \bibinfo{person}{Mottaqiallah Taouil}, \bibinfo{person}{Mohammad
  AlFailakawi}, {and} \bibinfo{person}{Said Hamdioui}.}
  \bibinfo{year}{2017}\natexlab{}.
\newblock \showarticletitle{Scouting logic: A novel memristor-based logic
  design for resistive computing}. In \bibinfo{booktitle}{\emph{ISVLSI'17}}.
  \bibinfo{pages}{176--181}.
\newblock


\bibitem[\protect\citeauthoryear{Xu, Dong, Jouppi, and Xie}{Xu
  et~al\mbox{.}}{2011}]%
        {DATE_2011_CongXu}
\bibfield{author}{\bibinfo{person}{Cong Xu}, \bibinfo{person}{Xiangyu Dong},
  \bibinfo{person}{Norman~P. Jouppi}, {and} \bibinfo{person}{Yuan Xie}.}
  \bibinfo{year}{2011}\natexlab{}.
\newblock \showarticletitle{Design implications of memristor-based RRAM
  cross-point structures}. In \bibinfo{booktitle}{\emph{2011 Design, Automation
  Test in Europe}}. \bibinfo{pages}{1--6}.
\newblock
\urldef\tempurl%
\url{https://doi.org/10.1109/DATE.2011.5763125}
\showDOI{\tempurl}


\bibitem[\protect\citeauthoryear{Zhang, Lin, Wang, Wang, Wang, Qian, and
  Huang}{Zhang et~al\mbox{.}}{2020}]%
        {Zhang_DATE2020}
\bibfield{author}{\bibinfo{person}{Yawen Zhang}, \bibinfo{person}{Sheng Lin},
  \bibinfo{person}{Runsheng Wang}, \bibinfo{person}{Yanzhi Wang},
  \bibinfo{person}{Yuan Wang}, \bibinfo{person}{Weikang Qian}, {and}
  \bibinfo{person}{Ru Huang}.} \bibinfo{year}{2020}\natexlab{}.
\newblock \showarticletitle{When Sorting Network Meets Parallel Bitstreams: A
  Fault-Tolerant Parallel Ternary Neural Network Accelerator based on
  Stochastic Computing}. In \bibinfo{booktitle}{\emph{2020 Design, Automation
  Test in Europe Conference Exhibition (DATE)}}. \bibinfo{pages}{1287--1290}.
\newblock
\urldef\tempurl%
\url{https://doi.org/10.23919/DATE48585.2020.9116390}
\showDOI{\tempurl}


\bibitem[\protect\citeauthoryear{Zidan, Strachan, and Lu}{Zidan
  et~al\mbox{.}}{2018}]%
        {Zidan2018}
\bibfield{author}{\bibinfo{person}{Mohammed~A. Zidan},
  \bibinfo{person}{John~Paul Strachan}, {and} \bibinfo{person}{Wei~D. Lu}.}
  \bibinfo{year}{2018}\natexlab{}.
\newblock \showarticletitle{The future of electronics based on memristive
  systems}.
\newblock \bibinfo{journal}{\emph{Nature electronics}}  \bibinfo{volume}{1}
  (\bibinfo{year}{2018}), \bibinfo{pages}{22--29}.
\newblock


\end{thebibliography}

\end{document}